\documentclass[prl, aps, amssymb, twocolumn, superscriptaddress, notitlepage, longbibliography]{revtex4-2}

\usepackage{amsmath}
\usepackage{amssymb}
\usepackage{amsthm}
\usepackage{amsfonts}
\usepackage{listings}
\usepackage{enumerate}
\usepackage{latexsym}
\usepackage{psfrag}
\usepackage{bm}
\usepackage{graphicx}
\usepackage[caption=false]{subfig}
\usepackage{blkarray}
\usepackage{array}
\usepackage{color}
\usepackage[dvipsnames]{xcolor}
\usepackage[normalem]{ulem}
\usepackage[colorlinks,linkcolor=blue,citecolor=blue,urlcolor=blue]{hyperref}
\usepackage{dsfont}
\usepackage{ulem}
\usepackage{bbold}
\usepackage{siunitx}
\usepackage{physics}
\usepackage{comment}
\usepackage{mathtools}
\usepackage{sidecap}
\usepackage{stmaryrd}
\usepackage{inputenc }

\def\R{{R}}

\definecolor{applegreen}{rgb}{0.55, 0.71, 0.0}

\newcommand{\beq}{\begin{equation}}
	\newcommand{\eeq}{\end{equation}}
\newcommand{\beqarray}{\begin{eqnarray}}
	\newcommand{\eeqarray}{\end{eqnarray}}

\newcommand{\fig}[1]{Fig.~\ref{#1}}

\newcommand{\pd}{{\phantom{\dag}}}

\usepackage{titlesec}
\newcommand{\beginsupplement}{%
        \setcounter{secnumdepth}{3}
        \setcounter{tocdepth}{3} 
        \setcounter{equation}{0}
        \renewcommand{\theequation}{S\arabic{equation}}%
        \setcounter{figure}{0}
        \renewcommand{\thefigure}{S\arabic{figure}}%
        \renewcommand{\thesubsection}{S\arabic{subsection}}
     }

\usepackage[format=plain,justification=centerlast]{caption}[2022/02/20]

\begin{document}
\title{Observation of non-Hermitian topology in a multi-terminal quantum Hall device}

\author{Kyrylo Ochkan}
\thanks{These two authors contributed equally}
\affiliation{Leibniz Institute for Solid State and Materials Research,
IFW Dresden, Helmholtzstrasse 20, 01069 Dresden, Germany}
\affiliation{W\"{u}rzburg-Dresden Cluster of Excellence ct.qmat, 01062 Dresden, Germany}

\author{Raghav Chaturvedi}
\thanks{These two authors contributed equally}
\affiliation{Leibniz Institute for Solid State and Materials Research,
IFW Dresden, Helmholtzstrasse 20, 01069 Dresden, Germany}
\affiliation{W\"{u}rzburg-Dresden Cluster of Excellence ct.qmat, 01062 Dresden, Germany}

\author{Viktor K\"{o}nye}
\affiliation{Leibniz Institute for Solid State and Materials Research,
IFW Dresden, Helmholtzstrasse 20, 01069 Dresden, Germany}
\affiliation{W\"{u}rzburg-Dresden Cluster of Excellence ct.qmat, 01062 Dresden, Germany}

\author{Louis Veyrat}
\affiliation{Leibniz Institute for Solid State and Materials Research,
IFW Dresden, Helmholtzstrasse 20, 01069 Dresden, Germany}
\affiliation{W\"{u}rzburg-Dresden Cluster of Excellence ct.qmat, 01062 Dresden, Germany}

\author{Romain Giraud}
\affiliation{Leibniz Institute for Solid State and Materials Research,
IFW Dresden, Helmholtzstrasse 20, 01069 Dresden, Germany}
\affiliation{Université Grenoble Alpes, CNRS, CEA, Grenoble-
INP, Spintec, F-38000 Grenoble, France}

\author{Dominique Mailly}
\affiliation{Centre de Nanosciences et de Nanotechnologies,
CNRS, Université Paris-Saclay, 91120 Palaiseau, France}

\author{Antonella Cavanna}
\affiliation{Centre de Nanosciences et de Nanotechnologies,
CNRS, Université Paris-Saclay, 91120 Palaiseau, France}

\author{Ulf Gennser}
\affiliation{Centre de Nanosciences et de Nanotechnologies,
CNRS, Université Paris-Saclay, 91120 Palaiseau, France}

\author{Ewelina M. Hankiewicz}
\affiliation{W\"{u}rzburg-Dresden Cluster of Excellence ct.qmat, 01062 Dresden, Germany}
\affiliation{Institute for Theoretical Physics and Astrophysics, Julius-Maximilians-Universit\"{a}t W\"{u}rzburg, D-97074 W\"{u}rzburg, Germany}

\author{Bernd B\"{u}chner}
\affiliation{Leibniz Institute for Solid State and Materials Research,
IFW Dresden, Helmholtzstrasse 20, 01069 Dresden, Germany}
\affiliation{W\"{u}rzburg-Dresden Cluster of Excellence ct.qmat, 01062 Dresden, Germany}
\affiliation{Department of Physics, TU Dresden, D-01062 Dresden, Germany}

\author{Jeroen van den Brink}
\affiliation{Leibniz Institute for Solid State and Materials Research,
IFW Dresden, Helmholtzstrasse 20, 01069 Dresden, Germany}
\affiliation{W\"{u}rzburg-Dresden Cluster of Excellence ct.qmat, 01062 Dresden, Germany}
\affiliation{Department of Physics, TU Dresden, D-01062 Dresden, Germany}

\author{Joseph Dufouleur}
\email{j.dufouleur@ifw-dresden.de}
\affiliation{Leibniz Institute for Solid State and Materials Research,
IFW Dresden, Helmholtzstrasse 20, 01069 Dresden, Germany}
\affiliation{W\"{u}rzburg-Dresden Cluster of Excellence ct.qmat, 01062 Dresden, Germany}

\author{Ion Cosma Fulga}
\email{i.c.fulga@ifw-dresden.de}
\affiliation{Leibniz Institute for Solid State and Materials Research,
IFW Dresden, Helmholtzstrasse 20, 01069 Dresden, Germany}
\affiliation{W\"{u}rzburg-Dresden Cluster of Excellence ct.qmat, 01062 Dresden, Germany}
	
\date{\today}

\begin{abstract}
Quantum devices characterized by non-Hermitian topology are predicted to show highly robust and potentially useful properties, but realizing them has remained a daunting experimental task.
This is because non-Hermiticity is often associated with gain and loss, which would require precise tailoring to produce the signatures of nontrivial topology.
Here, instead of gain/loss, we use the nonreciprocity of the quantum Hall edge states to directly observe non-Hermitian topology in a multi-terminal quantum Hall ring.
Our transport measurements evidence a robust, non-Hermitian skin effect: currents and voltages show an exponential profile, which persists also across Hall plateau transitions away from the regime of maximum non-reciprocity.
Our observation of non-Hermitian topology in a quantum device introduces a scalable experimental approach to construct and investigate generic non-Hermitian systems.
\end{abstract}

\maketitle

Hermitian quantum mechanics describes isolated quantum systems. These include conventional topological phases occurring in the ground states of certain materials \cite{Hasan2010}, such as the well-known quantum Hall phase, whose precisely quantized electrical resistance is used today in metrology \cite{Klitzing1980}.
When coupling a quantum device to the outside world, however, the resulting gain and loss lead to an effectively non-Hermitian description \cite{Ashida2020}.
Non-Hermitian systems can also be topologically nontrivial \cite{Bergholtz2021}, and thus have robust properties, some of which hold the promise for remarkable applications.
These include exponentially precise sensors \cite{Budich2020}, amplifiers \cite{Wang2022}, and light funnels \cite{Weidemann2020}.

Introducing gain and loss in a quantum device is easily achieved, for instance by not shielding it sufficiently well from its local environment.
However, customizing these two processes such as to reach a topological phase has so far remained challenging, and no quantum, condensed-matter devices showing non-Hermitian topology have been reported to date.
Instead, the existing experimental observations have been achieved using ultracold atoms \cite{Liang2022}, optical systems \cite{Xiao2020, Wang2021}, as well as using meta-materials governed by the laws of classical physics.
The latter include electronic circuits \cite{Helbig2020, Liu2021}, photonic crystals \cite{Weidemann2020}, as well as mechanical \cite{Brandenbourger2019, Ghatak2020} and acoustic \cite{Zhang2021a, Zhang2021b} systems.
Their operating principle is based on the fact that Kirchoff's laws, Newton's laws, and Maxwell's equations can be used to mimic the Schr\"{o}dinger equation describing the dynamics of quantum particles.

Here, we directly observe one of the characteristic signatures associated to non-Hermitian topology in the quantum regime of a condensed-matter system.
Rather than relying on gain and loss, we build on the quantum transport properties of a well-known, Hermitian topological phase: the quantum Hall phase.
Its unidirectional edge modes provide a link to non-Hermitian topology that can be accessed in conventional multi-terminal conductance measurements.
Our work introduces non-Hermitian topology and its potential applications to the field of experimental mesoscopic physics.
It opens the possibility to create devices 
that take advantage of
this type of topology, not just by using the quantum Hall effect, but also more generally by relying on the quantum transport properties of condensed-matter systems.

We start from one of the simplest examples of non-Hermitian topology -- the Hatano-Nelson (HN) model \cite{Hatano1996}. The Hamiltonian,
\begin{equation}\label{eq:HNHam}
{\cal H}^\pd_{\rm HN} = \sum_j J^\pd_{\rm left} c^\dag_{j-1} c^\pd_j + J^\pd_{\rm right} c^\dag_{j+1} c^\pd_j = \mathbf{c}^\dag H^\pd_{\rm HN} \mathbf{c},
\end{equation}
describes a chain on which quantum particles hop between neighboring sites (site index $j$, creation operator $c^\dag_j$), where the hopping to the left and to the right are real numbers with different magnitudes, $J_{\rm left} \neq J_{\rm right}$. 
$\mathbf{c}$ is a column vector formed from all the annihilation operators and $H_{\rm HN}$ is the Hamiltonian matrix, whose size is given by the number of sites in the chain.
	
This model is non-Hermitian, ${\cal H}^\pd_{\rm HN}\neq {\cal H}^\dag_{\rm HN}$ due to the nonreciprocal hoppings, and it is characterized by a net bulk current flowing in the direction of the larger hopping \cite{Zhang2020}.
Heuristically, all states in the bulk of the system will flow in the current direction, until eventually reaching the end of the chain. This is the boundary signature associated with its nontrivial topology, the non-Hermitian skin effect: in a finite chain with open boundary conditions, all eigenstates are exponentially localized at one end of the system (see Methods Sec.~\ref{SI:HN-model}).

Our work is based on the insight that, in the limit of maximal nonreciprocity, when the hopping in one direction vanishes identically, the HN model effectively describes a one-dimensional, unidirectionally propagating mode.
Thus, the Hamiltonian Eq.~\eqref{eq:HNHam}, even though it is non-Hermitian, provides an accurate description of the long-time dynamics of the quantum Hall edge \cite{Lee2019}.

\begin{figure}[ht]
	\includegraphics[width=.6\textwidth]{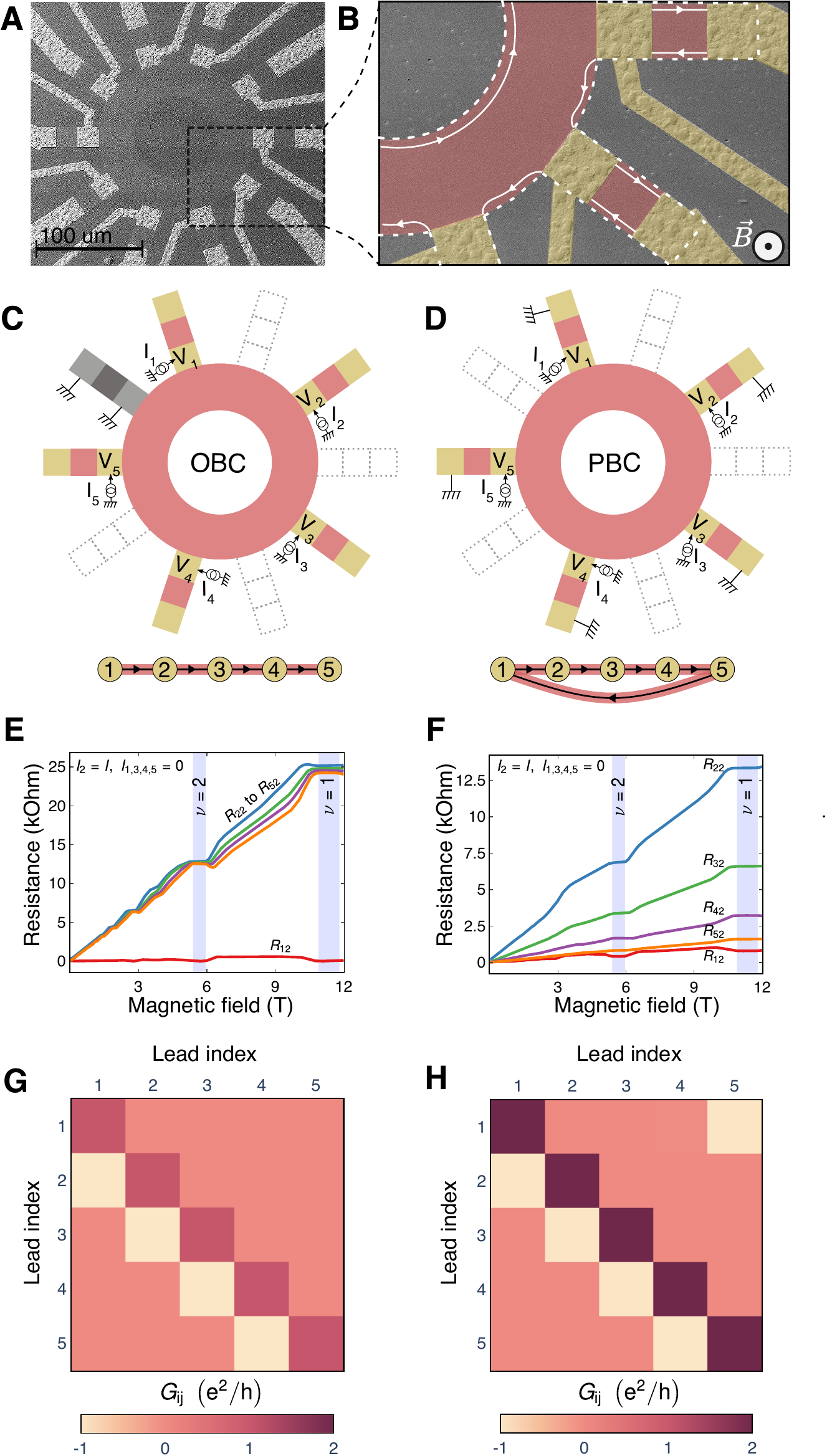}
	\captionlistentry{}
	\label{Fig. 1}

\caption*{
\noindent {\bf Fig. 1.} \textbf{Device schematic and characterization.} 
\textbf{(A)} Scanning electron microscopy (SEM) image of the AlGaAs 2DEG device. 
\textbf{(B)} Zoomed-in false-color SEM image. 
White lines indicate the edge quantum Hall states occurring in the presence of a perpendicular magnetic field at filling factor $\nu =1$ while white arrows indicate the direction of propagation of electrons. 
The 2DEG and ohmic contacts are highlighted in red and yellow, respectively. White dashed lines show the boundaries of the 2DEG.
\textbf{(C)} 5-site open-boundary condition (OBC) configuration where the grounded inactive arm (grey) is used to disconnect the first and the last site of the chain. 
The effective corresponding HN chain is shown schematically below. 
Active arms are labeled using currents and voltages, $I_j$ and $V_j$. 
For inactive arms in dashed lines, both contacts are floating.	
\textbf{(D)} 5-site periodic-boundary condition (PBC) configuration, with the effective HN chain shown schematically below.
\textbf{(E)} Measured magnetoresistance of all 5 sites of the OBC setup at $T=2.7$~K for current $I_2 = 10$~nA injected into the second site (the resistance matrix is shown explicitly in Methods Sec.~\ref{SI:measuring_resistance_matrix}).
\textbf{(F)} Measured magnetoresistance of the PBC setup for current  $I_2 = 10$~nA injected into the second site.
\textbf{(G)} Real part of the 5-site $G$ matrix measured at $B = 11.5$ T ($\nu=1$) for the OBC configuration in the same conditions as panel E.
\textbf{(H)} Real part of the 5-site $G$ matrix measured under the same experimental conditions for the PBC configuration in the same conditions as panel F.\hfill\break
}

\end{figure}

Starting from this insight, we have designed a multi-terminal quantum Hall device, shown in \fig{Fig. 1}A and B.
It consists of a two-dimensional electron gas (2DEG) ring etched in an AlGaAs/GaAs semiconducting heterostructure (see Methods Sec.~\ref{SI:characterization}), with arms distributed along its outer perimeter. 
Each arm consists of an `inner' ohmic contact, directly connected to the ring, and an `outer' contact, which is singly connected to the inner contact via a separate section of the 2DEG (see \fig{Fig. 1}B).
We label an arm as `active' when its inner contact is connected to a voltage probe and/or a current source.
The outer contacts of the active arms can be either grounded or floating.
In contrast, for an `inactive' arm, no current sources or voltage probes are attached to either contact. 
The contacts of an inactive arm can be either grounded 
or floating. 
There are a total of ten arms, but since the total number of measurements required scales quadratically with the number of active arms, not all of them are used.
Two examples of possible measurement configurations are shown in \fig{Fig. 1}C and D.
	
To highlight the connection between this device and non-Hermitian topology, we begin by considering the case of a single current source, which injects a current $I_j$ into the $j^{\rm th}$ active arm of the ring.
In the linear regime, the current $I_j$ is related to the voltages ${V_i}$ of the active arms as $I_j= \sum_{i} G_{ji} V_{i}$, where $G$ is the conductance matrix. In the presence of time-reversal symmetry, $G_{ij}=G_{ji}$, such that the conductance matrix is Hermitian. 
A magnetic field breaks this condition and induces non-Hermitian terms in $G$. 
Considering, for instance, the device in the quantum Hall regime at filling factor $\nu$ and in the presence of perfect contacts, the only nonzero elements of $G$ will be on its diagonal, as well as between adjacent contacts in the direction of propagation of the edge modes. 
For the $j^{\rm th}$ active arm (with $j>1$):
\begin{equation}
I_{j}+\nu\frac{e^2}{h} \left( V_{j-1}-\alpha V_{j}\right)=0,
\label{eq:Kirchhoff}
\end{equation}
with $\alpha=1$ if the outer contact of the arm is floating (see for instance \fig{Fig. 1}C) and $\alpha=2$ if the outer contact is grounded (see \fig{Fig. 1}D) and all inactive arms between $j-1$ and $j$ are floating. 
For the contact configurations shown in \fig{Fig. 1}C and D, the $G$ matrix is therefore related to the Hatano-Nelson Hamiltonian matrix with $J_\text{left}=0$ and $J_\text{right}=-1$:
\begin{equation}\label{eq:Gmat}
G = \nu\frac{e^2}{h} \left(H_{\rm HN} + \alpha \mathbb{1} \right),
\end{equation}
where $\mathbb{1}$ is the identity matrix.

The above equation lends itself to two different physical interpretations.
The first is that the quantum Hall ring is a meta-material. It is a way of generating a matrix, $G$, that is equivalent to the Hamiltonian matrix $H_{\rm HN}$ of a non-Hermitian quantum 
system, which 
would otherwise be out of experimental reach.
In this interpretation, the inner contact of each active arm of the quantum Hall ring plays the role of a site in the HN model, and the conductance matrix takes the role of the Hamiltonian.
In the second interpretation, Eq.~\eqref{eq:Gmat} means that due to the chiral nature of their edge modes, Chern insulators should show quantum transport properties whose presence and robustness are a consequence of non-Hermitian topology \cite{Franca2021}.
In the following, we shall explore the experimental consequences of both approaches.

Seen as a meta-material, the quantum Hall ring allows us to test two different configurations, equivalent to different realizations of the HN model.	
The first corresponds to open boundary conditions (OBC, \fig{Fig. 1}C), where the chain is cut between the last and the first site. 
This is achieved by grounding the inner contact of an intermediate inactive arm (between $j=N$ and $j=1$), hence setting $I_1=\alpha\nu e^2/h V_1$.
For periodic boundary conditions (PBC, \fig{Fig. 1}D), on the other hand, all inactive arms between $j=N$ and $j=1$ are floating, and the last site $N$ of the chain is connected to the first one: $I_1+\nu e^2/h (V_N-\alpha V_1)=0$. 
In the OBC configuration, the outer active contacts can be either grounded or floating
without affecting the topological properties of the chain, whereas in the PBC configuration, Kirchhoff’s laws require them to be grounded.

Experimentally, measuring the elements of $G$ will be achieved by first determining the elements of its inverse, the resistance matrix $R=G^{-1}$.
Injecting the current in each of the active arms, one by one, and measuring the voltages of all active arms using lock-in amplifiers yields each column of $R$ (see Methods Sec.~\ref{SI:measuring_resistance_matrix}).
The measurements of the second column of $R$ for any magnetic field and for the OBC and PBC configurations are shown in \fig{Fig. 1}E and \ref{Fig. 1}F, respectively. 

Permuting the position of the current source allows us to determine the full $R$ matrices. The corresponding conductance matrices $G$ are shown in \fig{Fig. 1}G and \fig{Fig. 1}H, respectively.
Owing to the robustness of the quantum Hall edge modes, the resulting, experimentally-measured conductance matrix at $\nu=1$ is remarkably close to that of a perfect, five-site HN chain, Eq.~\eqref{eq:Gmat} with $J_{\rm left}=0$, $J_{\rm right}=-1$, and $\alpha=2$.
We find that deviations in the individual matrix elements from the perfect values are of the order of a few percent or less.

\begin{figure}[ht]
	\includegraphics[width=\textwidth]{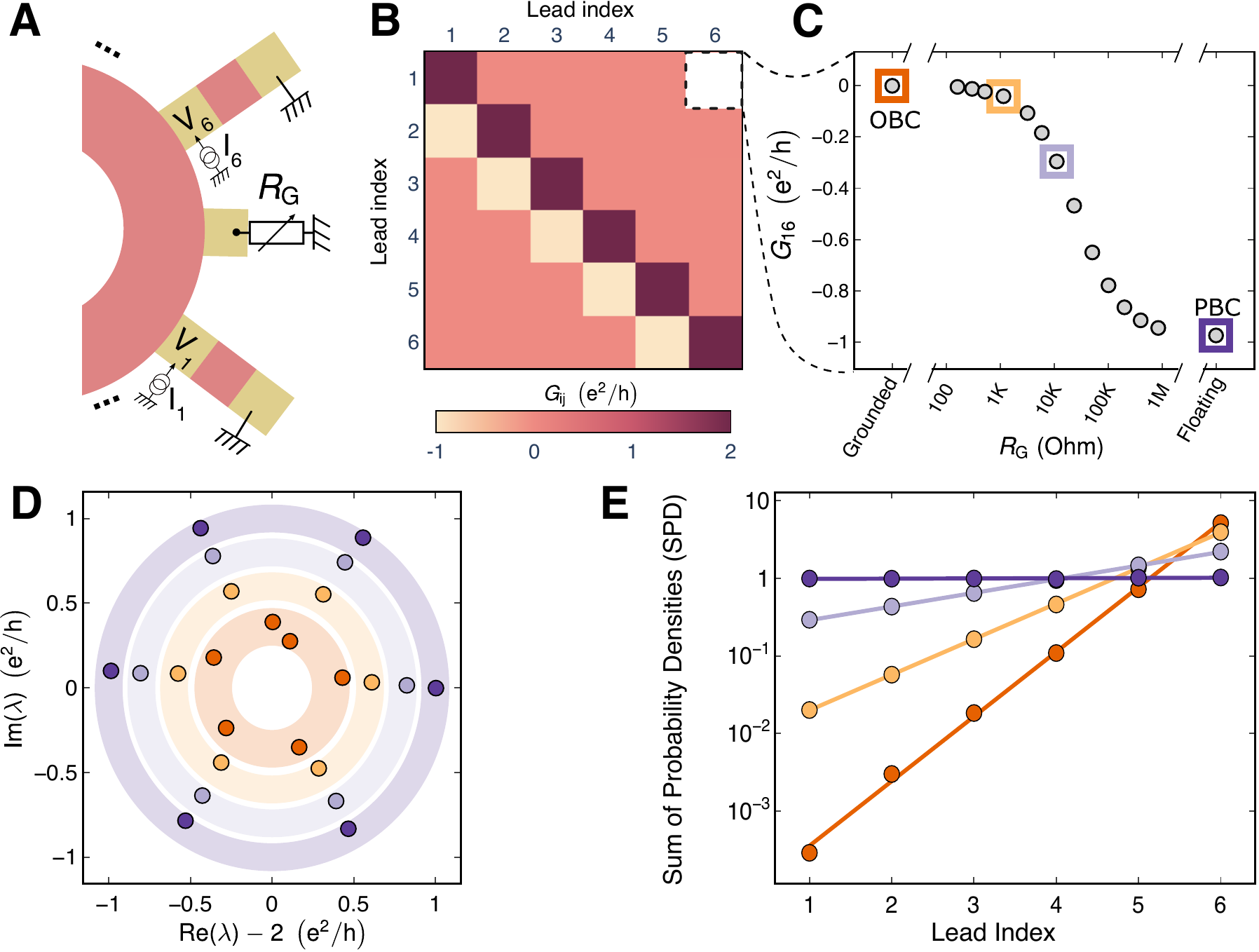}
	\captionlistentry{}
	\label{Fig. 2}

\caption*{
\noindent {\bf Fig. 2.} \textbf{Non-Hermitian skin effect.} 
\textbf{(A)} Schematic of the link between contact 1 and 6, including a variable resistance $R_{\mathrm{G}}$ at the end of the chain for continuous tuning between open and periodic boundary conditions. 
\textbf{(B)} Real part of the conductance matrix for the 6-site configuration at $B = 11.5$ T ($\nu = 1$). Changing the variable resistance $R_{\mathrm{G}}$ affects the top right element of the matrix (in white). 
This dependency is shown in panel C.
\textbf{(C)} Coupling term $G_{16}$ versus resistance $R_{\mathrm{G}} $ in logarithmic scale. The data points corresponding to the next panels are highlighted using squares.
\textbf{(D)} Eigenvalues of the experimental $G$ matrix (labeled $\lambda$) for the four selected points in units of $e^2/h$. 
The average diagonal element of the matrix is subtracted in order to center the plot around zero for better readability. 
We note that the eigenvalues do not collapse completely on a single point on the real axis. This is attributed to the finite resistance of the measurement line of the grounded contact (see Methods Sec.~\ref{SI:Additional_Coupling}).}
\textbf{(E)} Sum of probability densities (SPD) in logarithmic scale obtained by diagonalizing the measured $G$ matrix for the highlighted data points, plotted as a function of the lead index. 
The evolution from a constant to an exponentially decaying profile indicates the appearance of non-Hermitian skin effect upon gradually changing from PBC to OBC.  The markers indicate the experimental data points, and the lines are the best exponential fit to the data. \hfill\break

\end{figure}

Beyond the two configurations discussed above, the quantum Hall ring also allows us to continuously tune the effective HN chain from OBC to PBC. 
As shown in \fig{Fig. 2}A, for a six-site chain, this is achieved by connecting the inner contact of one of the inactive arms between $j=N$ and $j=1$ to the ground via a variable resistance $R_{\rm G}$, while its outer contact is floating. 
Changing the value of $R_{\rm G}$ serves to continuously tune the system between the grounded and floating configurations shown in \fig{Fig. 1}C and D and hence to tune it
from OBC to PBC (\fig{Fig. 2}B and \fig{Fig. 2}C).

As expected, numerically diagonalizing the $G$ matrix yields all the signatures associated with non-Hermitian topology in the HN model \cite{Okuma2020}. 
With PBC, its discrete eigenvalues are positioned along a circle in the complex plane, and gradually tuning towards OBC causes the eigenvalues to move inside of this circle (\fig{Fig. 2}D).
This is consistent with the theoretical prediction that all OBC eigenvalues are contained within the contour formed by the PBC spectrum (see also Methods Sec.~\ref{SI:HN-model}). 
At the same time, the probability density summed over all eigenvectors of the $G$ matrix (SPD, defined in Methods Sec.~\ref{SI:SPD_log_scale}) shows the non-Hermitian skin effect.
With PBC, this density is uniformly spread across the sites of the chain (meaning the arms of the ring), whereas moving towards OBC causes the probability density to become exponentially localized on the last site (\fig{Fig. 2}E, Methods Sec.~\ref{SI:SPD_log_scale} and \ref{SI:Additional_Coupling}), demonstrating the topologically non-trivial character of our device.

In the HN model, the presence and robustness of the non-Hermitian skin effect is predicted by a nonzero bulk invariant, similar to how the Chern number predicts the appearance of chiral edge modes.
For an infinite, translationally invariant chain, the non-Hermitian invariant is the winding number of the bulk bands as a function of momentum (see Methods Sec.~\ref{SI:HN-model}).
In our system, we have a finite number of sites, meaning that the bands are replaced by discrete eigenvalues.
In order to capture the non-Hermitian topology of the conductance matrix, and therefore the robustness of the skin effect, we determine topological invariants based on the methods recently developed for finite systems \cite{Hughes2021, Loring2015}. 
We use two distinct methods to determine the invariant, and thus to confirm that the non-Hermitian skin effect has a topological origin.
The two methods lead to invariants labeled as $w_{\rm PD}$ and $w_{\rm L}$. 
The $w_{\rm PD}$ invariant is based on the polar decomposition (PD) of the HN model, and was recently shown to correctly capture the topology of a finite-sized chain with OBC \cite{Hughes2021}.
It can take any real value, but approaches a quantized integer deep in the topological phase.
The $w_{\rm L}$ invariant, on the other hand, is integer by construction.
It is based on the so called `spectral localizer,' a mathematical object which has so far been used for the description of both Hermitian \cite{Loring2015} and non-Hermitian systems \cite{Liu2022, Cerjan2023}, and which we now apply to the HN model, again with OBC.
Both invariants are computed directly from the measured OBC conductance matrices, by using the formulas described in Methods Sec.~\ref{SI:topological_invariants}.

\begin{figure}[ht]
	\centering
	\includegraphics[width=.5\textwidth]{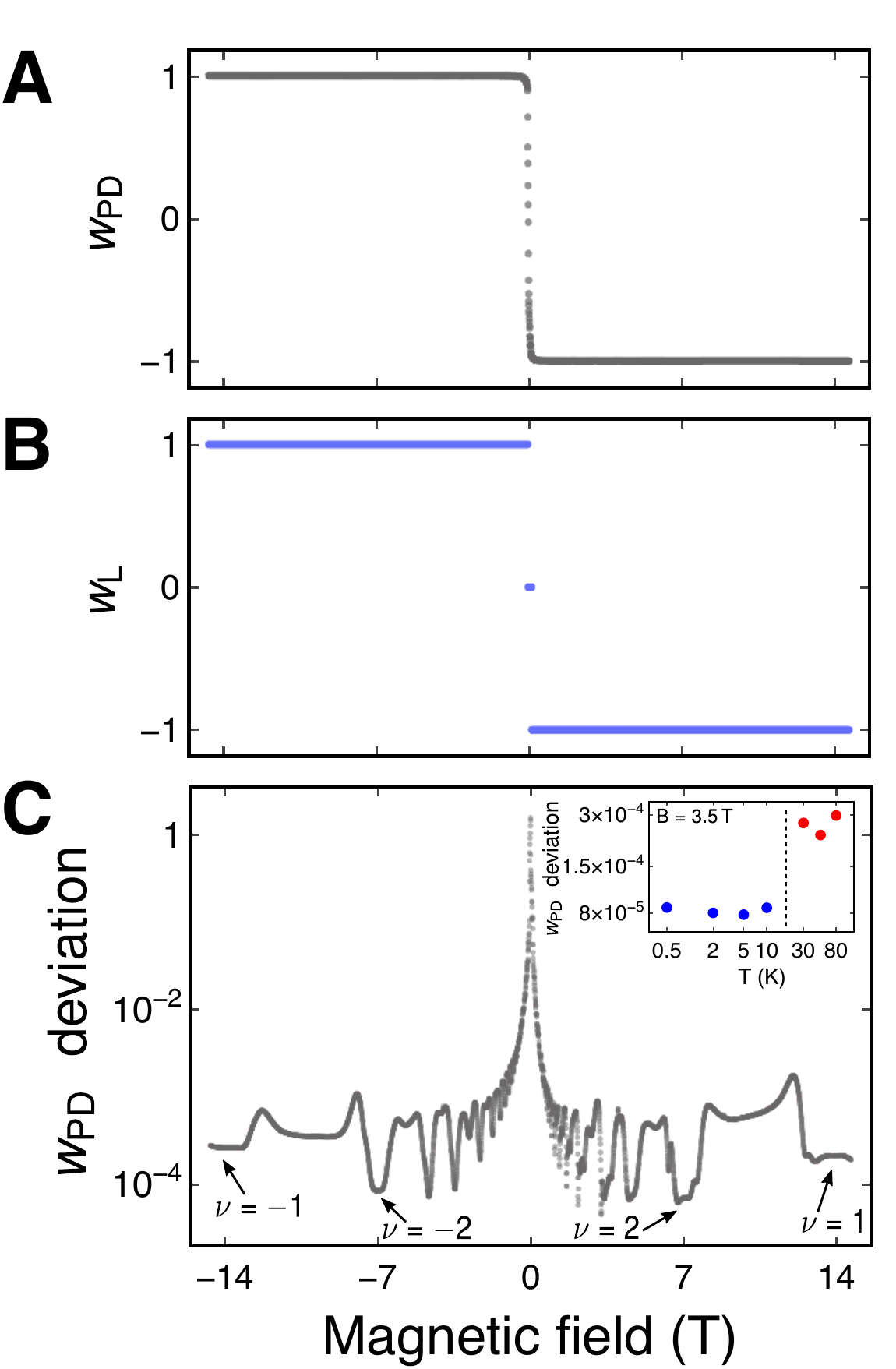}
	\captionlistentry{}
	\label{Fig. 3}

\caption*{
\noindent {\bf Fig. 3.} \textbf{Real-space topological invariants.} 
\textbf{(A)} Polar decomposition real-space topological invariant $\textit{w}_{\rm PD}$ and \textbf{(B)} localizer index $\textit{w}_{\rm L}$ for a 5-site OBC setup versus perpendicular magnetic field. 
\textbf{(C)} Deviation of the polar decomposition real-space invariant from the expected values ($-1$ for positive fields and $+1$ for negative fields), plotted on logarithmic scale versus magnetic field. 
The inset shows the invariant deviation measured at various temperatures in a QHE plateau at  $B = 3.5$~T, averaged over 100~mT.
The blue color denotes temperatures for which a quantum Hall plateau is measured whereas the red color denotes temperatures for which only dips (and no plateaus) are measured (see \fig{w_PD-Temp}). The vertical dashed line is a guide to the eye.}
\hfill\break

\end{figure}

As shown in \fig{Fig. 3}, the invariants reflect the nontrivial topology of the $G$ matrix at large magnetic fields. 
Depending on the direction of the magnetic field, and therefore of the chiral edge modes, both $w_{\rm PD}$ (\fig{Fig. 3}A) and $w_{\rm L}$ (\fig{Fig. 3}B) remain quantized to either $+1$ or $-1$, respectively.
Consistent with this, we find that the non-Hermitian skin effect is present either at the left or the right end of the chain (see Methods Sec.~\ref{SI:OBC_to_PBC_other_fields}), with all the eigenvectors positioned there.
Remarkably, the deviations of $w_{\rm PD}$ from the expected quantized value for a five-site chain remain on the order of $10^{-3}$ to $10^{-4}$ for a wide range of magnetic fields, including across plateau transitions (see \fig{Fig. 3}C).
Consistent with this, the non-Hermitian skin effect is present also for magnetic fields at which the Hall conductance is not quantized (see Methods Sec.~\ref{SI:OBC_to_PBC_other_fields}).
Thus, we find that the non-Hermitian invariant is more robust than the Chern number.
This is the case, for instance, when the magnetic field sets the position of the Fermi energy on a Landau level, or when the formation of Landau levels is prevented by disorder - at very low mobility and high temperature or at low magnetic fields - and therefore when the 
Chern number is not defined (see Methods Sec.~\ref{SI:DiffTempAndDensity}). The best quantization of $w_{\rm PD}$ improves to $10^{-5}$ for an eight-site chain and for Fermi energy set between two Landau levels. It should be improved further for a larger number of sites until it will be limited by the coupling between the last ($j=N$) and the first ($j=1$) sites, the coupling to previous sites (backscattering) or the noise level of the measurements (see Methods Sec.~\ref{SI:invariants_for_different_leads} and \ref{SI:Additional_Coupling}).
Finally, we note that the field asymmetry observed in \fig{Fig. 3}C is a result of onsite disorder in the HN chain, as discussed in Methods Sec.~\ref{SI:JumpsAndAsymmetry}. 

We now discuss the transition between the classical diffusive regime and the quantum-Hall regime
with maximal non-reciprocity. In the classical regime, as well as between the quantum Hall plateaus, Eqs.~\eqref{eq:Kirchhoff} and \eqref{eq:Gmat} are not valid anymore and finite long-range couplings (between distant sites) appear in the conductance matrix, 
similar
to the conductance matrix at zero magnetic field. The emergence of finite off-diagonal elements in the upper triangle of the conductance matrix leads to a less robust quantization of $w_\text{PD}$ and a broader skin effect (see Methods Sec.~\ref{SI:Additional_Coupling}). We tune the amplitude of the off-diagonal elements by varying on one hand the magnetic field, and on the other hand the mobility
through changing
the temperature.
At small magnetic fields, $|B|\lesssim\SI{0.4}{\tesla}$, both the invariants and the skin effect lose their robustness (see Methods Sec.~\ref{SI:OBC_to_PBC_other_fields}).
$w_{\rm PD}$ becomes poorly quantized (\fig{Fig. 3}C), whereas $w_{\rm L}$, which is integer by definition, becomes equal to $0$ or begins to rapidly oscillate between the values $-1$, $0$, and $+1$.
This behavior is a typical signature of topological phase transitions occurring in finite-sized systems (here, a five-site chain).

A similar loss of robustness is observed as a function of increasing temperature.
In order to tune the onset of the quantum regime, the conductance matrix was measured at different temperatures between 470 mK and 80K, thus changing
the mobility of the device over a broad range. The topological invariants are calculated at all fields and temperatures (Methods Sec.~\ref{SI:DiffTempAndDensity}).
To highlight the importance of quantum effects, we show in the inset of \fig{Fig. 3}C the temperature dependence of the invariant quantization at a fixed magnetic field.
When the temperature is lowered and the system enters a quantum Hall plateau regime, the invariant quantization shows a sharp improvement, by a factor between 3 and 4.
We have confirmed that the best quantization is systematically obtained for magnetic fields and temperatures that set the system well on a QHE plateau.

Surprisingly, for small fields, the field dependence of the invariant quantization is roughly temperature and therefore mobility independent. 
It scales quadratically with the field, even far into the classical regime (Methods Sec.~\ref{SI:DiffTempAndDensity}). 
In contrast, two measurements done with two different electronic densities indicate that the electron density strongly influences the field dependence of the invariant quantization: the lower the density, the sharper the decay (see Methods Sec.~\ref{SI:DiffTempAndDensity}). 
We note that unexpected features are brought to light by the results at the highest temperatures ($T=50$ K and 80 K),
where the evolution of the topological invariants shows jumps that are discussed in more details in Methods Sec.~\ref{SI:DiffTempAndDensity} and \ref{SI:JumpsAndAsymmetry}.

In analogy to the methods commonly used in meta-materials such as topoelectric circuits, we have so far measured the $R=G^{-1}$ matrix element by element using a single current source, and then examined topological features by diagonalizing the $G$ matrix numerically.
In contrast, the conventional quantum Hall effect, meaning the robust quantization of the Hall conductance, is a directly-observable physical phenomenon, independent of any effective model.

\begin{figure}[ht]
	\centering
	\includegraphics[width=\textwidth]{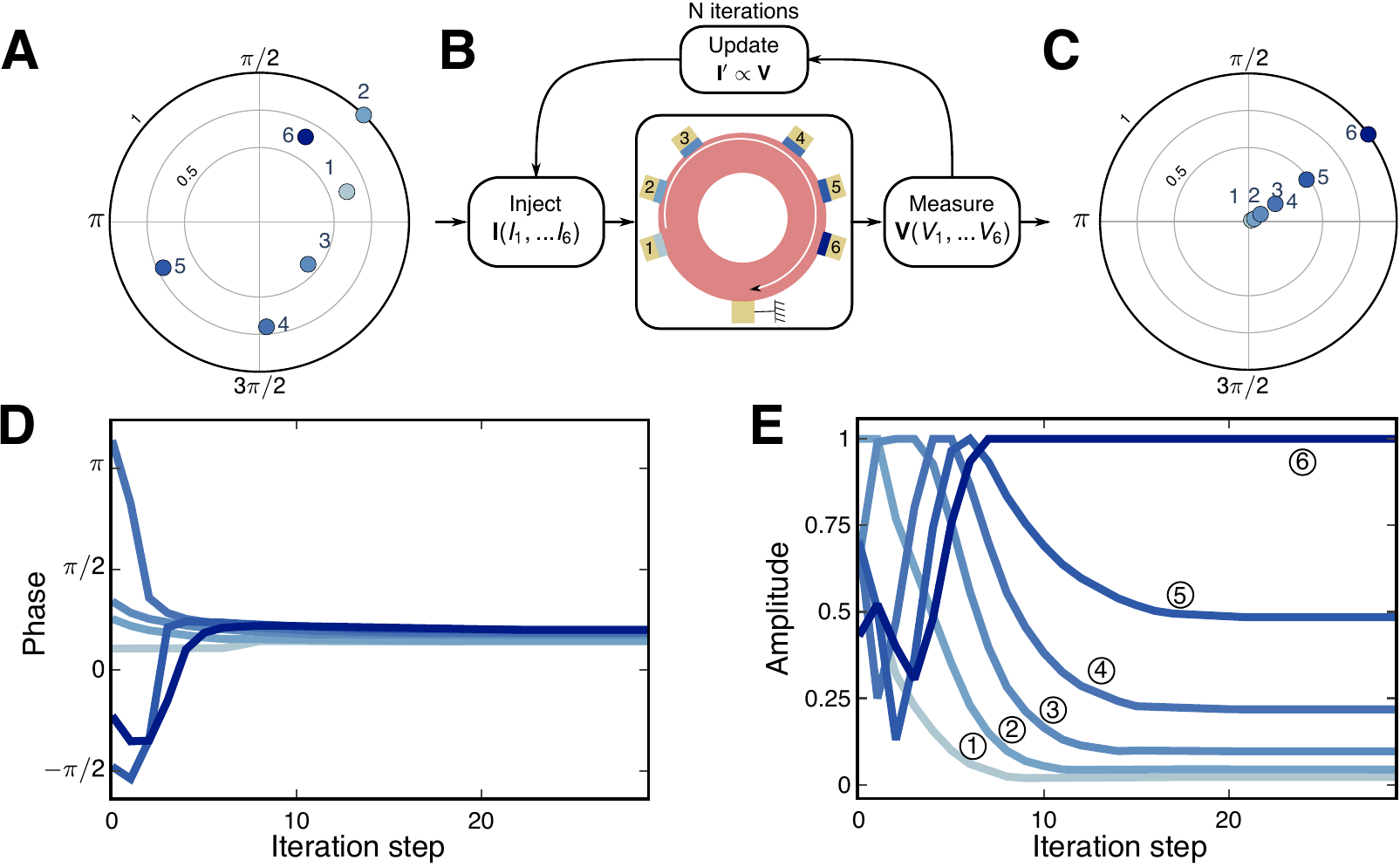}
	\captionlistentry{}
	\label{Fig. 4}

\caption*{
\noindent {\bf Fig. 4.} \textbf{Measurement of the non-Hermitian skin effect via iteration for the OBC setup.} 
\textbf{(A)} Elements of the initial, randomly generated current vector (see Methods Sec.~\ref{SI:iteration_explanation}) displayed in polar coordinates for a 6-site set-up. 
The amplitude is normalized such that the different coordinates are in units of the largest injected current ($150$~nA). The measurement is performed at $B = 5.5$~T ($\nu=2$)
\textbf{(B)} Flowchart of the iterative procedure. 
\textbf{(C)} Final current configuration in the system after 40 iterations. 
\textbf{(D)} Evolution of the phase of each vector element versus iteration number. 
The system converges towards a final current configuration with a unique phase as a function of iteration number (see also Methods Sec.~\ref{SI:iteration_explanation}).
\textbf{(E)} Evolution of the amplitude of each element versus iteration number, in units of the largest injected current ($150$~nA). 
The final current configuration is reached after 15 to 20 iterations, when the amplitudes no longer change as a function of iteration number. This final current configuration shows an exponential decay as a function of lead index, from 6 to 1 (from dark to light blue), which is a direct manifestation of the non-Hermitian skin effect in experiment (see also Methods Sec.~\ref{SI:iteration_explanation}). \hfill\break
}

\end{figure}

Turning to the second interpretation of Eq.~\eqref{eq:Gmat}, we now show that, similar to
the quantized Hall conductance of the QHE, the topological, non-Hermitian skin effect is a directly-observable transport property of our device, which does not require the determination of the $G$ matrix nor its numerical diagonalization.
To achieve this, we make use of the fact that the skin effect implies that conductance matrix eigenvectors are all exponentially localized at one boundary of the system. 
One general numerical method to determine an eigenvector is the so-called iterative power method. 
By applying iteratively a matrix $A$ to a random non-zero vector $\mathbf{x}$, one converges toward the largest eigenvalue $\lambda$ of $A$: 
\begin{align}
\lim_{n \rightarrow \infty} A^n \mathbf{x} = \lambda^n \mathbf{v},
\end{align} 
with $\mathbf{v}$ the eigenvector of $A$ associated to $\lambda$. 
This can be experimentally realised in our device by simultaneously applying different random currents to all active contacts using multiple current sources (so having a random current vector $\mathbf{I_0} = (I_1, \ldots,I_N)$, see \fig{Fig. 4}A), followed by measuring the resulting voltages $V_j$ of each site $j$, and then iteratively re-applying currents to each contact such that the new current vector is proportional to the previous voltage vector $\mathbf{I_{n+1}} \propto \mathbf{V_n}$. 
This is effectively applying $G^n$ to the current vector $\mathbf{I_0}$, as pictured in \fig{Fig. 4}B. 
Repeating this iteration should drive voltages to converge towards one of the eigenvectors of $G$, all of which show the non-Hermitian skin effect.

We applied this scheme to our quantum Hall device with six active contacts, labeled 1 to 6. 
After 15 to 20 such iterations, we observe that both the current and the voltage vectors become independent of the iteration step (\fig{Fig. 4}C). 
After having converged, both vectors show an exponentially decaying profile over the contacts from 6 to 1 (\fig{Fig. 4}D and \fig{Fig. 4}E).
This constitutes a direct measurement of the non-Hermitian skin effect: 
the current vector converges to one of the eigenvectors of the conductance matrix (see Methods Sec.~\ref{SI:iteration_explanation}), all of which are exponentially localized to one end of the Hatano-Nelson chain (here contact 6). 
The exponential nature of the current and voltage profiles is a robust transport signature of our quantum device. It does not change for a wide range of magnetic fields, it is independent of the initially chosen currents, and it is only present in the OBC configuration (see Methods Sec.~\ref{SI:iteration_explanation}).
The robustness of this multiple-source transport signature is a direct consequence of non-Hermitian topology in our quantum Hall ring.

\paragraph{\bf Author contributions} JD and ICF conceived and supervised the project. JD and KO designed the sample, conducted the measurements and analysed the data with input from LV and RG. DM fabricated the sample on GaAs/AlGaAs heterostructures grown by AC and UG. RC performed numerical simulations of the Hatano-Nelson model, computed the polar-decomposition invariant, and developed the localizer index, under the supervision of VK, EH, JvdB, and ICF. VK introduced the idea of the iterative measurement, and performed numerical simulations showing its feasibility. All authors participated to interpreting the results and writing of the manuscript.

\paragraph{\bf Data availability} The data and codes used in this work are available on Zenodo at \cite{zenodocode}.

\paragraph{\bf Acknowledgements} We thank Ulrike Nitzsche for technical assistance. JD would like to thank his departed friend and colleague Fabien Portier for introducing him to the physics of the quantum Hall effect.
This work was supported by the French RENATEC network, by DIM NANO-K, and by the Deutsche Forschungsgemeinschaft (DFG, German Research Foundation) under Germany's Excellence Strategy through the W\"{u}rzburg-Dresden Cluster of Excellence on Complexity and Topology in Quantum Matter -- \emph{ct.qmat} (EXC 2147, project-ids 390858490 and 392019).
LV was supported by the Leibniz Association through the Leibniz Competition.

\let\oldaddcontentsline\addcontentsline
\renewcommand{\addcontentsline}[3]{}
\bibliography{article.bib}
\let\addcontentsline\oldaddcontentsline
	
\clearpage 
\onecolumngrid 


\clearpage

\clearpage

\clearpage

\clearpage


\beginsupplement
\section*{Methods}

\tableofcontents

\subsection{Hatano-Nelson Model} \label{SI:HN-model}

The Hatano-Nelson (HN) model is the simplest example of a non-Hermitian system with nonreciprocal hoppings. 
Defined on a one-dimensional lattice with no symmetry requirements, this model can be described by a single-orbital tight-binding Hamiltonian \cite{Hatano1996}:

\begin{equation}\label{equation-Hatano_Nelson_Hamiltonian}
{\cal H}^\pd_{\rm HN} = \sum_j J^\pd_{\rm right} c^\dag_{j+1} c^\pd_j + J^\pd_{\rm left} c^\dag_{j-1} c^\pd_j = \mathbf{c}^\dag H^\pd_{\rm HN} \mathbf{c},
\end{equation}
where $J^\pd_{\rm left}$ and $J^\pd_{\rm right}$ $\in \R $ are unequal hopping amplitudes. 
For a finite-sized chain in real-space, the Hamiltonian matrix takes the form
\begin{equation}\label{eq:SI_HNmatrix}
H_{\rm HN} = 
    \begin{pmatrix}
      0 & J_{\rm left} & & & & \\
      J_{\rm right} & 0 & J_{\rm left} & & & \\
      & J_{\rm right} & 0 & \ddots & & \\
      & & \ddots & \ddots & \ddots & \\
      & & & \ddots & 0 & J_{\rm left} \\
      & & & & J_{\rm right} & 0 \\
    \end{pmatrix}.
\end{equation}
If the chain is infinitely long, Fourier transforming yields the one-band Bloch Hamiltonian:
\begin{equation} \label{equation-Hatano_Nelson_Hamiltonian-k}
    H(k) = J^\pd_{\rm right} e^{-ik} + J^\pd_{\rm left} e^{+ik}.
\end{equation}
Its eigenvalues form an ellipse in the complex plane and wind around the origin. 
The corresponding non-Hermitian winding number $w$ takes the general form \cite{Gong2018}
\begin{equation} \label{equation-nonhermitian winding number}
w(E_B) = \frac{1}{2 \pi i} \int_{0}^{2 \pi} \partial_k \log \det \left (  H(k) - E_B  \right ) dk,
\end{equation}
where $E_B$ is the so-called base point, i.e., the point in the complex plane at which the winding number is evaluated.
This invariant characterizes the non-Hermitian topology of the system, manifesting itself in the non-Hermitian skin effect.

Calculating Eq.~\eqref{equation-nonhermitian winding number} for $H(k)$ at $E_B = 0$ gives:
\begin{equation}
w =
  \begin{dcases*}
     -1 \quad |J^\pd_{\rm right}| > |J^\pd_{\rm left}|;\\
    +1  \quad |J^\pd_{\rm right}| < |J^\pd_{\rm left}|.
  \end{dcases*}
\end{equation}

\begin{figure}[tbh]
	\centering
	\includegraphics[width=16.44cm]{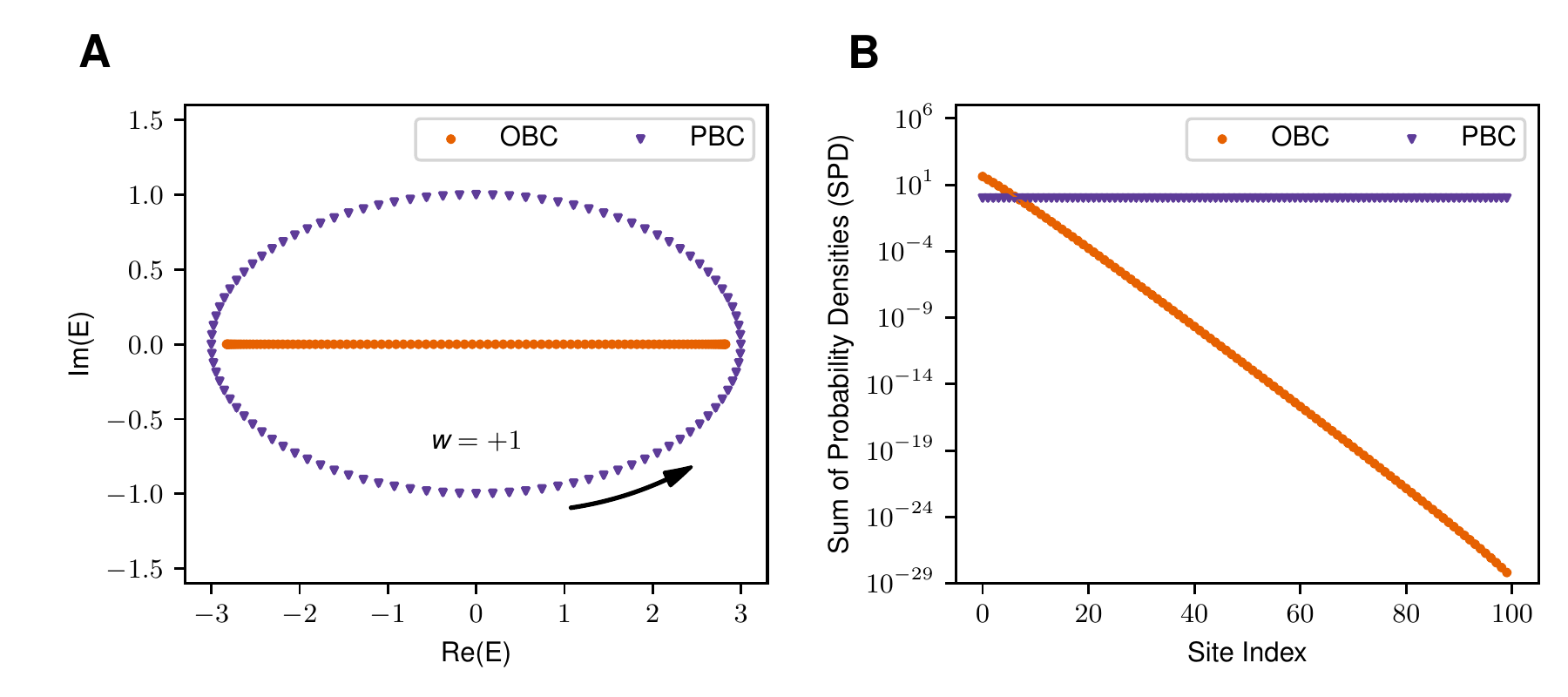}
	\captionlistentry{}
	\label{Fig. SI. 3.1}
    
    \caption*{
	\noindent {\bf Fig. S1.} \textbf{Eigenspectra and skin effect. } 
	\textbf{(A)} Eigenspectra for OBC and PBC. The eigenvalues are real for OBC, and complex for PBC. 
	The arrow indicates the spectral winding expected in the Bloch Hamiltonian, with winding number $w= +1$.
	\textbf{(B)} SPD plotted against the sites of the system on a log-linear scale under OBC and PBC; We use $(J^\pd_{\rm left},J^\pd_{\rm right}) = (2,1)$ and a chain consisting of 100 sites. \hfill\break
	}
\end{figure}

The eigenvalues and eigenvectors of $ {\cal H}^\pd_{\rm HN}$ are sensitive to the boundary conditions imposed on the system. Under open boundary conditions (OBC), the eigenvalues are real (see Fig.~\ref{Fig. SI. 3.1}A).
Defining the sum of probability densities (SPD) at the $j^{th}$ lattice site as $\sum_i | \bra{r_j} \ket{\Psi_i} |^2$, where $\ket{\Psi_i}$ are the right eigenvectors of ${\cal H}^\pd_{\rm HN}$ and $\ket{r_j}$ denote the lattice site positions, we observe that an extensive number of eigenstates are localized at the ends of the system (see OBC configuration in Fig.~\ref{Fig. SI. 3.1}B), a phenomenon unique to non-Hermitian systems and commonly referred to as the non-Hermitian skin effect \cite{Yao2018}. 
Moreover, this localization occurs at the right (left) end of the system for $|J^\pd_{\rm right}|/|J^\pd_{\rm left}| > 1(<1)$, and is related to the nontrivial winding of the energy bands \cite{Okuma2020}. 
However, with periodic boundary conditions (PBC), the spectrum consists of discrete points positioned on an ellipse (see Fig.~\ref{Fig. SI. 3.1}A), thus reproducing the behavior of the Bloch Hamiltonian characterizing the infinite system, Eq.~\eqref{equation-Hatano_Nelson_Hamiltonian-k}. 
The system's translational invariance leads to a constant SPD ($=1$) on all sites (see Fig.~\ref{Fig. SI. 3.1}B).

When approaching the limit of maximum nonreciprocity, $J_{\rm left} = 0$ and $J_{\rm right} = 1$, the PBC spectrum evolves towards a circle in the complex plane, whereas the OBC spectrum moves towards the origin, $E=0$.
A gradual transition between PBC and OBC then involves a shrinking of the initially circular spectrum: eigenvalues progressively move inside of the contour defined by the PBC spectrum, as can be seen for the experimentally determined $G$ matrix in \fig{Fig. 2}.
Further, slight deviations in the matrix elements away from the perfect limit of Eq.~\eqref{eq:SI_HNmatrix}, which in experiment are of the order of few percent or less, lead to slight changes in the positions of individual eigenvalues.

\subsection{Device characterization} \label{SI:characterization}

\begin{figure}[tbh]
	\centering
	\includegraphics[width=.6\textwidth]{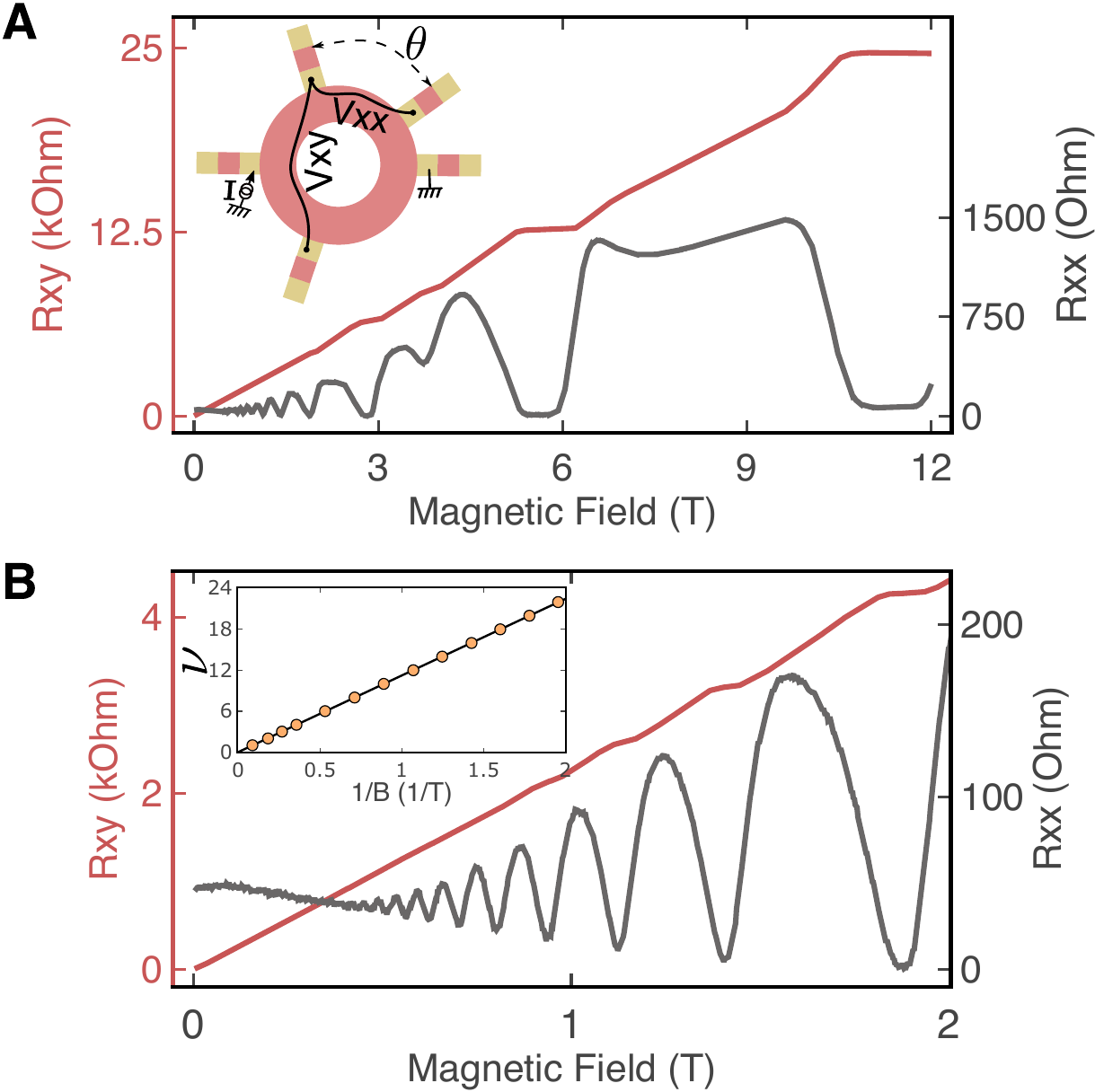}
	\captionlistentry{}
	\label{fig:SI:characterization}

    \caption*{
	\noindent {\bf Fig. S2.} \textbf{Sample characterization. } 
	\textbf{(A)} Longitudinal and transverse magnetoresistance. Inset shows the measurement schematic.
	\textbf{(B)} Zoom-in of the panel A. 
	The inset shows the filling number versus the coordinates of the respective valleys in the Shubnikov–de Haas oscillations.\hfill\break
	}
\end{figure}

The transport properties of the device at temperature $T\sim 2.1$~K were measured similarly to those of standard two-dimensional electron gas Hall bars. 
The connection schematic is shown in the inset of panel A in \fig{fig:SI:characterization} and a magnetic field $B$ is applied in the out-of-plane direction. 
A DC current of $50$~nA was injected in the source contact, whereas the opposite contact is grounded and used as a drain. 
All other contacts are floating and some are used as voltage probes. 
The transverse and longitudinal voltages are measured in a four-probe configuration and the magnetic field dependence of the associated resistances (respectively $R_\mathrm{xx}$ and $R_\mathrm{xy}$) is shown in \fig{fig:SI:characterization}A with a zoom in at low fields in \fig{fig:SI:characterization}B.

The electron density $n$ is extracted from both the transverse magnetoresistance $R_\mathrm{xy}(B)$ and the Shubnikov-de Haas oscillations in $R_\mathrm{xx}(B)$. 
At low magnetic fields, the classical Hall effect is directly related to the electron density by the relation $R_\mathrm{xy} = -B/ne$, with $e$ the electronic charge. 
A linear fit of the magnetic field dependence of the Hall resistance gives $n=2.75 \times 10^{11}\,\text{cm}^{-2}$.

Additionally, the electron density can be estimated from the analysis of Shubnikov-de Haas oscillations. 
At low fields, each maximum of $R_\mathrm{xx}(B)$ corresponds to a Landau level crossing the Fermi energy, or, equivalently, the minima of $R_\mathrm{xx}(B)$ correspond to a Fermi energy lying between two Landau levels. 
At high magnetic field, Zeeman spin-splitting occurs and the Landau levels become spin nondegenerate. 
In order to avoid any issues induced by the spin-splitting, we plot in the inset of \fig{fig:SI:characterization}B the position of the minima of $R_\mathrm{xx}(B)$ as a function of $1/B$ and we label them with an index $\nu$ that stands for the number of Landau levels lying below the Fermi energy, taking into account the spin degeneracy. 
Hence, the first minimum measured at $B\sim 11$~T corresponds to $\nu=1$ and the second minimum at $5.5$~T corresponds to $\nu=2$. 
For fields $B<3$~T, the spin-splitting induced minima disappear and the index takes only even values. 
Therefore, the index $\nu$ coincides with the filling factor and indicates how many edge channels participate to charge transport.
The charge density can again be extracted thanks to the relation $\nu=n h/eB$ where $h$ is the Planck constant. 
We find $n=2.72 \times 10^{11}\,\text{cm}^{-2}$, in very good agreement with the value extracted from Hall measurements.

In addition, the two-probe magnetoresistance of each arm of the device was measured to probe the homogeneity of the sample. 
We observed small variations of about 2\% in resistance at $\nu=1$, indicating weak inhomogeneities of the sample and slightly different quality of the ohmic contacts.
However, this does not affect the overall results in any substantial way.

In order to estimate the mobility of the sample, we use the expression that relates the four-probe resistance $R$ of the device at zero magnetic field to the mobility $\mu$ in a ring geometry:
\begin{equation}\label{equation-mobility}
    R=\frac{1}{2}\frac{ \theta}{ne\mu \ln{\left(R_2/R_1\right)}},   
\end{equation}

with $\theta$ the angle between the two contacts in radians, $R_1$ the radius of the inner circle, $\R_2$ the one of the outer circle and the factor $1/2$ accounting for the two arms of the ring that makes the effective current in a single arm half of the total current. 
We have $\mu = 5.03 \times 10^{5}\,\text{cm}^{2} / \text{Vs}$, in good agreement with the nominal mobility of the 2DEG ($\mu = 6.6 \times 10^{5}\,\text{cm}^{2} / \text{Vs}$ measured on van der Pauw samples). This corresponds to a transport length $\ell = 4.3 \mu$m, much smaller than the width of the ring or the distance between two adjacent contacts ($\approx 32.5 \mu$m). It sets the system in the diffusive regime at low magnetic fields.

 During a second set of measurements, 
 the charge density and the mobility was extracted at different temperatures, ranging from 470 mK to 80 K.
 Due to a faster cool-down, the electron density of the 2DEG 
 at 1.7 K was lower by about 20~\% compared to the first set of measurements.
The mobility changes by roughly a factor~$5$ between 470~mK and $80$~K, whereas the charge density remains almost temperature independent. The different densities and mobilities for the different temperatures can be found in Table~\ref{Table:SI:density_mobility}.

 	\begin{table}[h!]
			\begin{tabular}{|c|c|c|c|c|}
                \hline
				\rule[-2ex]{0pt}{0pt}
                \rule{0pt}{3ex}
				\hspace{1mm} Temperature \hspace{1mm} & \hspace{1mm} SdH density (cm$^{-2}$) \hspace{1mm} & \hspace{1mm} Hall density (cm$^{-2}$) \hspace{1mm} & \hspace{1mm} Hall mobility (cm$^{2}$.V$^{-1}$.s$^{-1}$) \hspace{1mm} & \hspace{1mm} transport length $\ell$ ($\mu$m) \hspace{1mm} \\
				\hline
				\rule{0pt}{3ex}
				470~mK & 3.32 $\times$ 10$^{11}$ & 3.54 $\times$ 10$^{11}$ & 5.4 $\times$ 10$^{5}$ & 5.3 $\mu$m \\
				1.7~K & 3.30 $\times$ 10$^{11}$ & 3.67 $\times$ 10$^{11}$ & 5.1 $\times$ 10$^{5}$ & 5.1 $\mu$m \\
				5~K & 3.35 $\times$ 10$^{11}$ & 3.56 $\times$ 10$^{11}$ & 5.1 $\times$ 10$^{5}$ & 5.0 $\mu$m \\
				10~K & 3.31 $\times$ 10$^{11}$ & 3.58 $\times$ 10$^{11}$ & 4.8 $\times$ 10$^{5}$ & 4.7 $\mu$m \\
                30~K &  & 3.33 $\times$ 10$^{11}$ & 3.3 $\times$ 10$^{5}$ & 3.1 $\mu$m \\
                50~K &  & 3.35 $\times$ 10$^{11}$ & 2.5 $\times$ 10$^{5}$ & 2.4 $\mu$m \\
				\rule[-2ex]{0pt}{0pt}
                80~K &  & 3.58 $\times$ 10$^{11}$ & 1.2 $\times$ 10$^{5}$ & 1.2 $\mu$m \\
                \hline
			\end{tabular}
		\caption{Densities and mobilities corresponding to the second run of measurements, at temperature between 470~mK and 80~K. As explained in the text, the densities are obtained from either the Shubnikov-de Haas oscillations (second column) or the Hall voltage in the $5$~contact configuration (third column). The forth column corresponds to the calculation of the Hall mobility using Eq.~\eqref{equation-mobility} and the Hall charge density. In the last column, we calculate the transport length $\ell$ associated with the Hall density and the Hall mobility.}
        \label{Table:SI:density_mobility}
	\end{table}
 
We used the $5$-site OBC configuration to measure and calculate the parameters in Table~\ref{Table:SI:density_mobility} in a similar way
as
for the first run of measurements (see Inset in Fig. \ref{SI:characterization}). At low enough temperatures, the charge density extracted from Shubnikov-de Haas oscillations is in fair agreement with Hall results. The low temperature Hall mobility, calculated from Eq. \eqref{equation-mobility}, led to a similar mobility
as
the one obtained in the first set of measurements with lower charge density.

\subsection{Measurement of the resistance matrix and determination of the conductance matrix} \label{SI:measuring_resistance_matrix}

The conductance matrix $G$ is determined by measuring the resistance matrix $R$ and calculating the inverse of the resistance matrix using $G=R^{-1}$. 
The column $j$ of the $R$~matrix is measured by injecting a current $I_j$ into the inner contact $j$ of the sample and measuring the voltage $V_{i}$ on each inner contact $i$.
The grounding of outer and inner contacts depends on which configuration (OBC or PBC) is measured, and is done according to the scheme shown in \fig{Fig. 1}C and D in the main text. 
The different elements $R_{ij}$ of the $R$~matrix are given by $R_{ij}=V_{i}/I_{j}$.

\begin{figure}[tbh]
	\centering
	\includegraphics[width=1\textwidth]{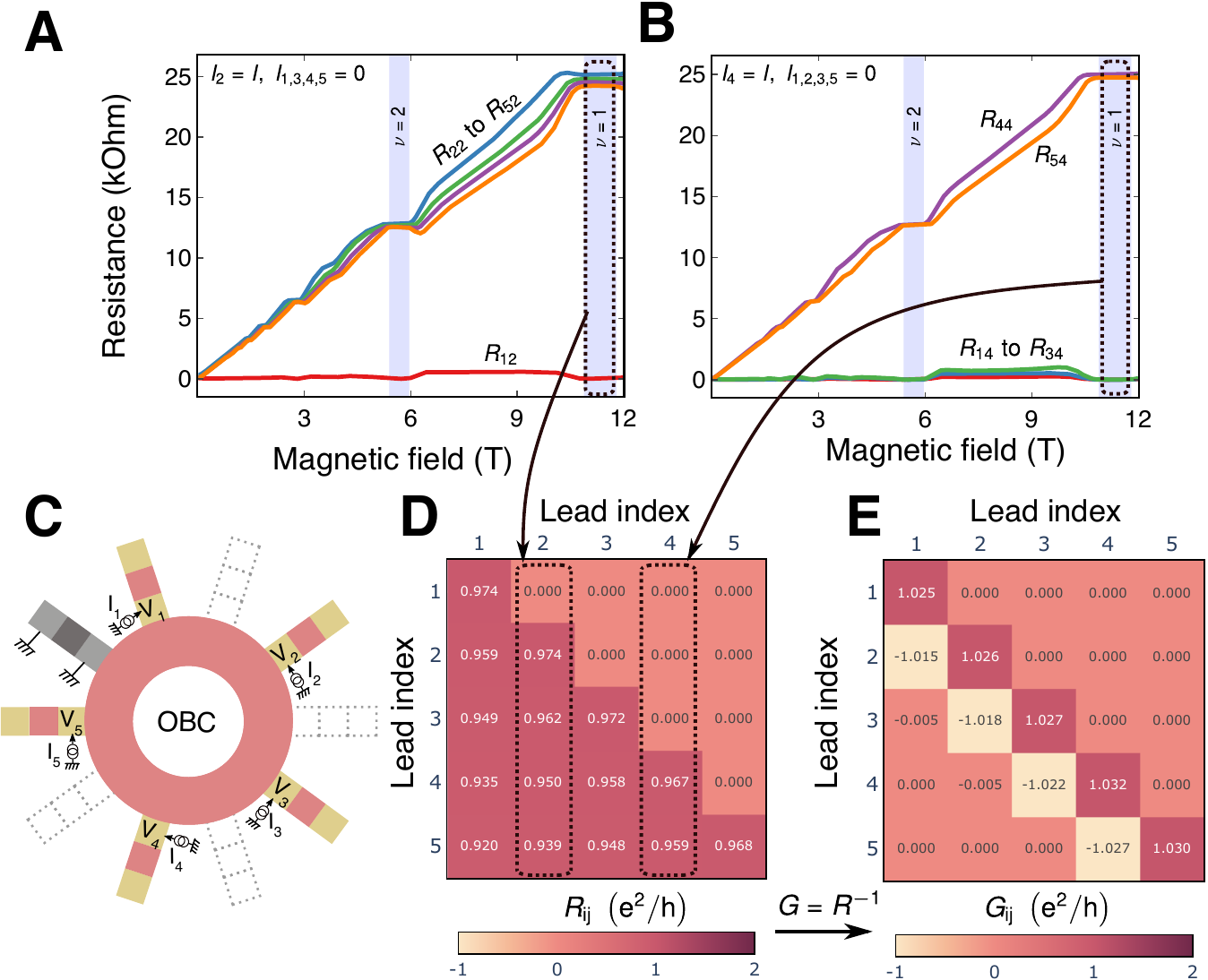}
	\captionlistentry{}
	\label{fig:SI:resistance_matrix}
	
	\caption*{
    \noindent \noindent {\bf Fig. S3} \textbf{Resistance and conductance matrix measurement procedure.} 
    \textbf{(A)} Measured magnetoresistance of the OBC setup for current  $I_2 = 10$~nA injected into the second site at $T=2.1$~K.
	\textbf{(B)} Measured magnetoresistance of the 5-lead OBC setup for current  $I_4 = 10$~nA injected into the fourth site at $T=2.1$~K.
	\textbf{(C)} Experimental 5-lead OBC setup.
	\textbf{(D)} Measured resistance matrix $R$ at $B = 11.5$~T ($\nu = 1$). The second and fourth columns that are determined from the panel (A) and (B), respectively, are highlighted. 
	\textbf{(E)} Obtained conductance matrix constructed by inverting the resistance matrix R. 
	\hfill\break
	}
	
\end{figure}

By sweeping the magnetic field while injecting current into one of the contacts, we measure the magnetic field dependence of the $R$~matrix and determine $G(B)$, as seen in \fig{fig:SI:resistance_matrix}. 
The measurement is done by sweeping the magnetic field at a given field sweep rate while sampling voltages at time intervals $\Delta t$. 
The magnetic field dependence of each element at a given magnetic field is interpolated in order to obtain the accurate dependence of $R(B)$. 
Special care was taken to adjust the field sweep rate such that the final number of interpolated points remains lower than the original number of experimental data points. 

\begin{figure}[tbh]
	\centering
	\includegraphics[width=1\textwidth]{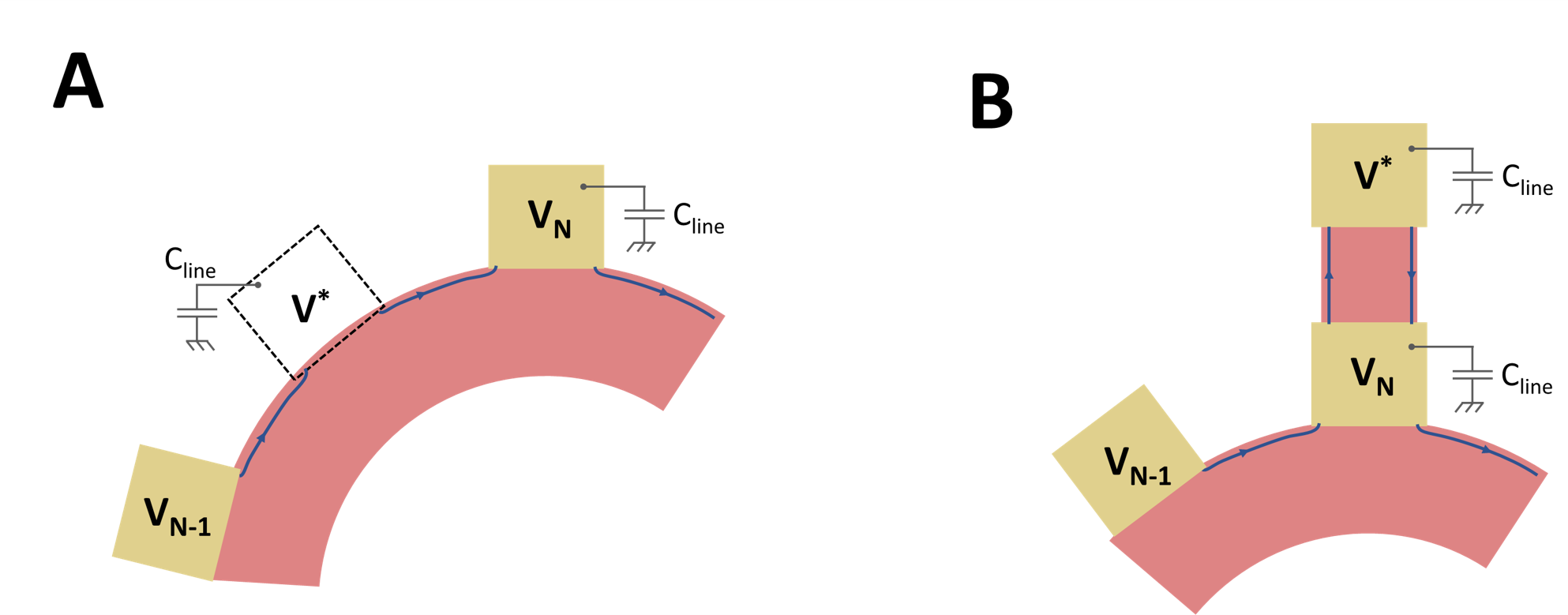}
	\captionlistentry{}
	\label{fig:SI:Capa}

    \caption*{
	\noindent {\bf Fig. S4.} \textbf{Electronic schematics including the capacitance of the measurement lines. } 
	\textbf{(A)} Configuration including a floating contact (dashed line) between two adjacent active contacts. The contact is connected to the ground via the capacitance of the measurement line.
	\textbf{(B)} Configuration including a floating outer contact connected to the ground via the capacitance of the measurement line.\hfill\break
	}
\end{figure}

Experimentally, we apply an AC current of 10~nA at 
$\omega/2\pi=173.2$~Hz, and we measure the voltage using standard lock-in preamplifier techniques. 
The measured voltages have in-phase and out-of-phase components. 
The large capacitance of the measurement lines (about $1$~nF per line) implies that the out-of-phase signal is not always negligible with respect to the in-phase signal. 
More quantitatively, in order to determine both the in-phase and out-of phase voltages of the different contacts, one has to consider the capacitive connection to the ground for each floating or voltage probe contact. 
We detail below two generic configurations that lead to a contribution of the capacitance to the out-of-phase signal.
The case of a floating contact lying between two adjacent contacts is shown in \fig{fig:SI:Capa}.A and the case of a floating outer contact is shown in \fig{fig:SI:Capa}.B.

For both cases exposed in \fig{fig:SI:Capa}, the voltage $V_\textrm{N}$ is related to $V_\textrm{N-1}$ to the first order in $C_\textrm{line}\omega h\backslash \nu e^2$ by $V_\textrm{N}=\left( 1-2jC_\textrm{line}\omega h\backslash \nu e^2\right) V_\textrm{N-1}$, 
where we only assumed $C_\textrm{line}\omega h\backslash \nu e^2 \ll 1$. More generally, since every contact is connected to a capacitive measurement line (even when floating), each contact contributes to the total capacitance such that 
\begin{equation}
    V_\textrm{N}=\left( 1-jM\frac{h}{\nu e^2}C_\textrm{line}\omega\right) V_\textrm{N-1},
    \label{SI:Eq:Capa}
\end{equation}
where $M$ is the number of floating contacts between the contacts $N-1$ and $N$ (the contact $N$ being included). 
Generally, the measured out-of-phase signal is in very good agreement with Eq.~\eqref{SI:Eq:Capa}. Depending on the configuration, the out-of-phase signal can be up to 50~\% of the in-phase signal.

In contrast to the out-of-phase signal, the correction to the in-phase signal by the capacitive measurement lines is of second order in $C_\textrm{line}\omega h\backslash \nu e^2$. Nevertheless, the large capacitance leads to some deviations, as observed in the \fig{Fig. 1} of the main text.

\subsection{Experimental determination of the SPD} \label{SI:SPD_log_scale}

In our experimental set-up, determining the sum of probability densities (SPD), as defined in Methods Sec.~\ref{SI:HN-model}, 
involves the components of the $G$ matrix eigenvectors. 
Labeling $V^{j}_{i}$ the $i^\text{th}$ component (related to the site $i$ of the chain) of the $j^\text{th}$ normalized eigenvector $\textbf{V}^{j}$ of the experimentally determined $G$ matrix, the SPD reads
\begin{equation}
  \text{SPD}(i) = \sum_j |V^{j}_{i} |^2
\end{equation}
where SPD($i$) is the SPD at the site $i$.

\subsection{Impact of additional coupling terms in $G$} \label{SI:Additional_Coupling}

In order to understand the impact of different coupling terms related to both the grounding of the chain between the first and last contact in the OBC, and the impact of the diffusive regime at low magnetic fields, we investigate here the consequences of having some finite element in the diagonal $i$ of the G matrix, with $i=2,...,N$ or $i=-3,...,-N$. Positive $i$ corresponds to the upper diagonal terms and negative $i$ to the lower diagonal terms. Hence, $i=2$ corresponds to the coupling between a contact and the previous one in the matrix~\eqref{eq:SI_GWithCoupling}, and $i=N$ corresponds to the coupling between the last site ($i=N$) and the first site ($i=1$).

\begin{equation}\label{eq:SI_GWithCoupling}
G = \nu \frac{e^2}{h}
    \begin{pmatrix}
      \alpha & 0 & & \varepsilon & & \\
      -1 & \alpha & 0 & & \ddots & \\
      & -1 & \alpha & \ddots & & \enspace \varepsilon \\
      & & \ddots & \ddots & \ddots & \\
      & & & \ddots & \alpha & \enspace 0\\
      & & & & -1 & \enspace \alpha \\
    \end{pmatrix}.
\end{equation}

It is possible to show that, generally, a long-range coupling to next-nearest neighbour in the forward direction ($i<0$) does not have any impact neither on the module of the eigenvalues of $G$ nor on its eigenstates, whereas for $i>0$ (back-scattering type of coupling), the impact of a small coupling $\varepsilon$  is very strong on the eigenvalues $\lambda$, which become proportional to $\varepsilon^{1/i}$ (for instance, $\lambda= \sqrt[i]{(N-i+1)\varepsilon}$ for $i \leq (N+1)/2$). Moreover, for $\varepsilon \ll 1$, the eigenstates of $G$ are exponentially localized at the end of the chain with $a_\text{j} \propto \lambda^\text{j}$, $a_\text{j}$ being the $j$th component of the eigenvector.

Therefore, the modules of the eigenvalues is a power function of the coupling between the last and the first site of the chain and shows an exponential sensitivity with the number of site ($\lambda \propto \varepsilon^{1/N}$). As explained in the main text, the decoupling of the last and first site is obtained by grounding an ohmic contact between the two sites. Such a contact is grounded at room temperature on the measurement box through measurement lines, the resistance of which is much smaller than the quantum of resistance (about two orders of magnitude smaller). Nevertheless, this finite resistance is at the origin of the coupling between the last and first sites. Because of the exponential sensitivity of the system to this term, such a finite coupling has a strong impact on the eigenvalues of the OBC system even in the infinitesimal limit, and the ellipse, constituted by the eigenvalues of $G$ in OBC, does not collapse to a single point at the center of the PBC eigenvalues as expected for $\varepsilon=0$. Considering the coupling measured in our OBC setup, we expect a shrinkage of the eigenvalues ellipse by about 50\% (instead of 100\%) for 5 or 6 sites 
in agreement with the experimental measurements
(see \fig{Fig. 2} in the main text and \fig{fig:SI:SPD_other_fields} below). 

\begin{figure}[tbh]
	\centering
	\includegraphics[width=1\textwidth]{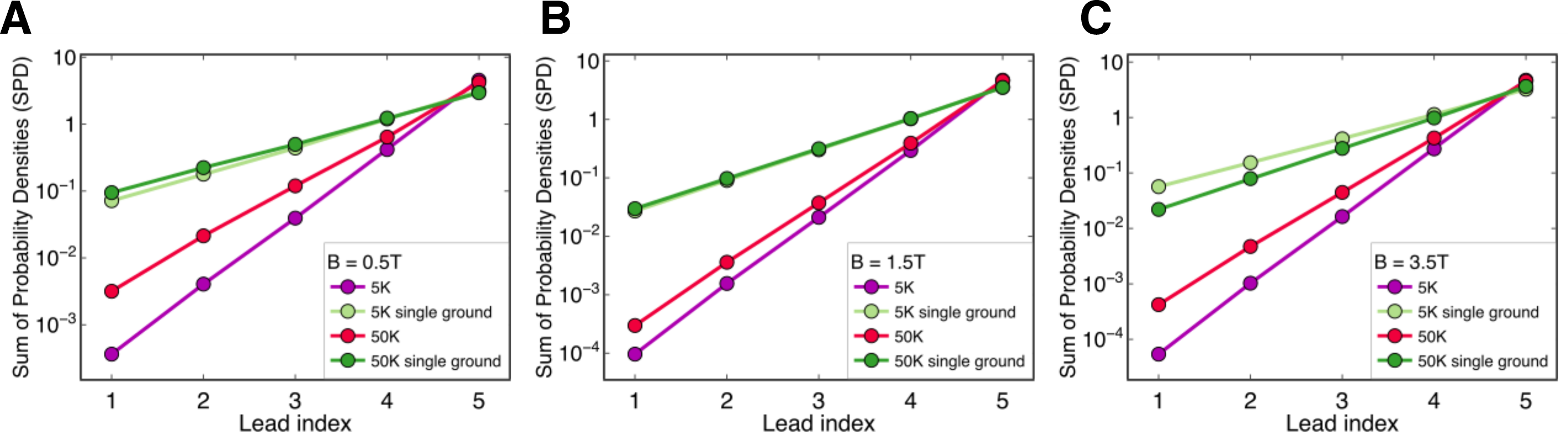}
	\captionlistentry{}
	\label{fig:SI:SPD_ground}
    
    \caption*{
	\noindent {\bf Fig. S5.} \textbf{Comparing the SPD for the setup with a single and multiple grounded contacts at $5$~K and $50$~K at} \textbf{(A)} $B= 0.5$~T, \textbf{(B)} $B= 1.5$~T,  \textbf{(C)} $B= 3.5$~T.  \hfill\break
	}
\end{figure}

Similarly to the impact on the spectrum of $G$, the coupling between the last and the first 
site
strongly affects the skin effect, which has been investigated experimentally by having two different grounding 
schemes
associated to two different couplings. In a first experiment, we cut the chain by grounding a single contact, while in a second experiment, the chain is 
interrupted
by three grounded contacts in series, all of them being grounded at room temperature via resistive measurement lines (about 130~Ohms). In the quantum Hall effect regime, the finite resistance of the measurement lines implies that the grounded ohmic contact does not absorb all the current flowing from the last contact to the first one. Instead, a small but finite current is emitted from the grounded contact to the first contact of the chain between the last and first site. Introducing additional contacts 
allows us to
reduce this leak current exponentially with the number of grounded contacts. Even at low magnetic fields, much below the onset of the quantum Hall effect regime, the coupling term between the last and the first site greatly affects the skin effect as seen in \fig{fig:SI:SPD_ground}.

The impact of further coupling terms in the $G$ matrix are discussed below for skin effect at low magnetic field and temperature (see Section \ref{SI:OBC_to_PBC_other_fields}).

\subsection{Real-space topological invariants} \label{SI:topological_invariants} 

A recent work has proposed a real-space formula to calculate the topological invariant of finite-sized Hatano-Nelson chains \cite{Hughes2021}. 
This invariant, which we label $w_{\rm PD}$, uses the polar decomposition of the Hamiltonian matrix at base point $E_B$,
\begin{equation}\label{TL_Hughes_Invariant_Equation}
    H_{\rm HN} - E_B \mathbb{1} = Q P,
\end{equation}
where $Q$ is unitary, and $P$ is a positive-definite matrix. $w_{\rm PD}$ takes the form:
\begin{equation} \label{equation-TH-02}
w_{\rm PD} = \mathcal{T}(Q^{\dag}[Q, X]),
\end{equation}
where $X={\rm diag}(1, 2, \ldots, L)$ encodes the positions of the sites of the Hatano-Nelson chain and $\mathcal{T}$ is the trace per unit volume evaluated over a middle interval of the chain, $[l+1,L-l]$. 

Since the conductance matrix $G$ shown in the main text describes an effective 1D Hatano-Nelson model, we replace $H_{\rm HN}$ in Eq.~\eqref{TL_Hughes_Invariant_Equation} by $G$ to determine the associated invariant. For the 6-site setup
($L = 6$)
we have chosen $l = 2$, and $E_B$ is the arithmetic mean of the diagonal entries of the conductance matrix $G$.
In general, for a long enough chain the choice of $l$ is not expected to qualitatively change the value of $w_{\rm PD}$, provided that the trace of Eq.~\eqref{equation-TH-02} is performed over sites sufficiently far from the boundaries of the chain \cite{Hughes2021}.

The second invariant, $w_{\rm L}$, is based on the so-called `spectral localizer,' a mathematical object recently used to derive topological invariants for finite-sized Hermitian systems \cite{Loring2015, Loring2019, Michala_2021}.
We adapt this invariant to the Hatano-Nelson model by using the fact that non-Hermitian topology is related to Hermitian topology by doubling the Hamiltonian \cite{Okuma2020}. 
In our case, the Hatano-Nelson Hamiltonian matrix is related to the well-known Su-Schrieffer-Heeger \cite{Su1979} (SSH) model as,
\begin{equation}\label{eq:SI_HN_to_SSH}
    H_{\rm SSH} =
    \begin{pmatrix}
      0 & H^\pd_{\rm HN} - E^\pd_B \\
      H^\dag_{\rm HN} - E^*_B & 0 \\
    \end{pmatrix}.
\end{equation}

It has been shown that the topological invariant of $H_{\rm SSH}$ is equal to that of $H_{\rm HN}$ evaluated at base point $E_B$ \cite{Okuma2020}.
Thus, we use Eq.~\eqref{eq:SI_HN_to_SSH} together with the localizer index of the SSH chain \cite{Loring2015}, resulting in:
\begin{equation}\label{eq:SI_localizer_index_SSH}
    w_{\rm L} = \frac{1}{2} {\rm sig}
    \begin{pmatrix}
      X & -i H^\pd_{\rm HN} +i E^\pd_B \\
      i H^\dag_{\rm HN} - i E^*_B & -X \\
    \end{pmatrix}.
\end{equation}
Here, ${\rm sig}$ denotes the matrix signature, i.e., the number of positive eigenvalues minus the number of negative eigenvalues, and $X$ labels the sites in the chain, now with the condition that the origin is positioned close to the middle of the system: 
$X = {\rm diag} (0,1,2, \ldots, L-2, L-1) - \lfloor L/2 \rfloor \mathbb{1}$, where $ \lfloor x \rfloor$ is the integer part of x.
As before, we replace the Hatano-Nelson Hamiltonian matrix with the measured conductance matrix, $G$, and set $E_B$ to be the mean of its diagonal elements.

\subsection{Non-Hermitian skin effect for various magnetic fields, temperatures and boundary conditions} \label{SI:OBC_to_PBC_other_fields}

\begin{figure}[tbh!]
	\centering
	\includegraphics[width=.5\textwidth]{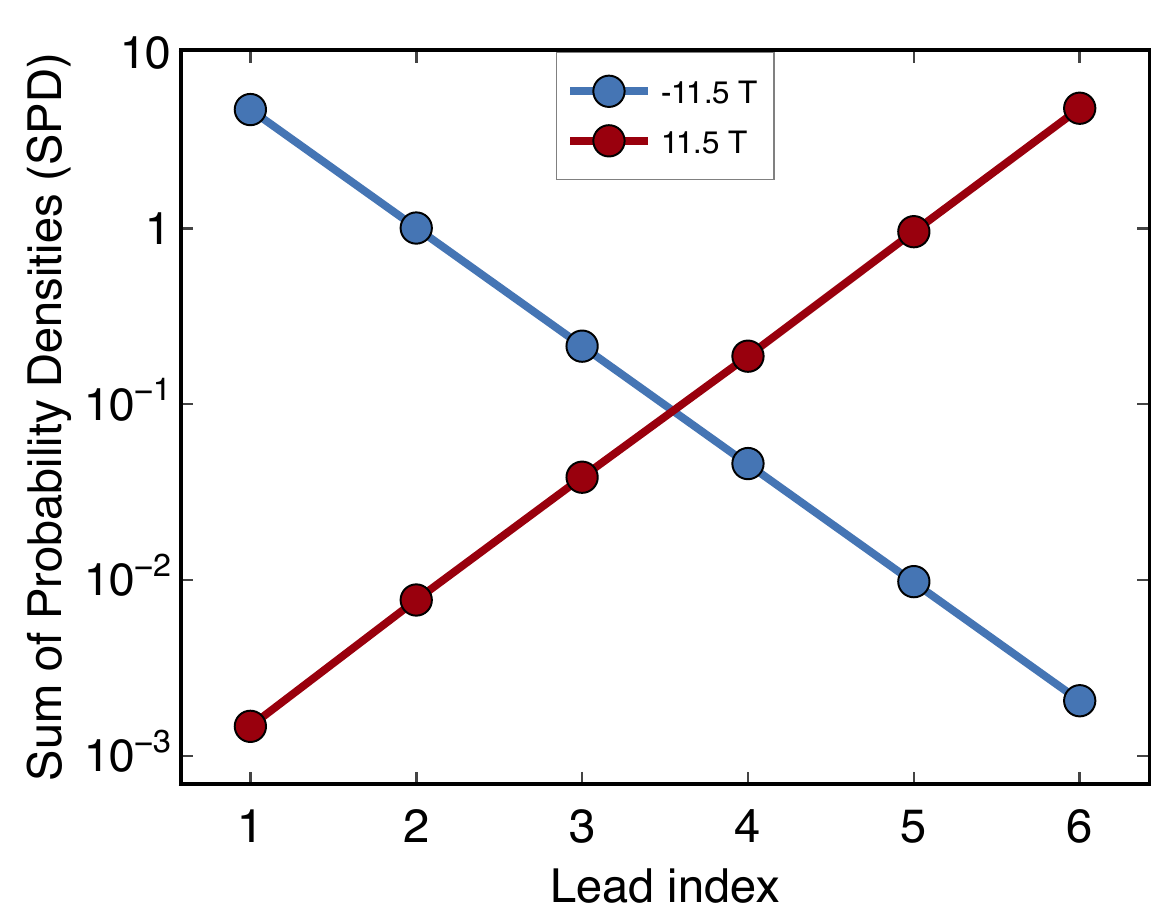}
	\captionlistentry{}
	\label{fig:SI:SPD_negative_field}

    \caption*{
	\noindent {\bf Fig. S6.} \textbf{Comparing the SPD for positive and negative external magnetic field.} 
	The magnetic fields correspond to Chern numbers of $\nu=1$ and $\nu=-1$. \hfill\break
	}
\end{figure}

We show in \fig{fig:SI:SPD_negative_field} the experimental non-Hermitian skin effect for both positive and negative magnetic fields.
The field strength is chosen such that there is a single chiral edge mode at the Fermi level ($\nu=1$), whose direction depends on the magnetic field orientation.
The two resulting skin effects are mirror symmetric with respect to the middle of the Hatano-Nelson chain.

\begin{figure}[tbh]
	\centering
	\includegraphics[width=.8\textwidth]{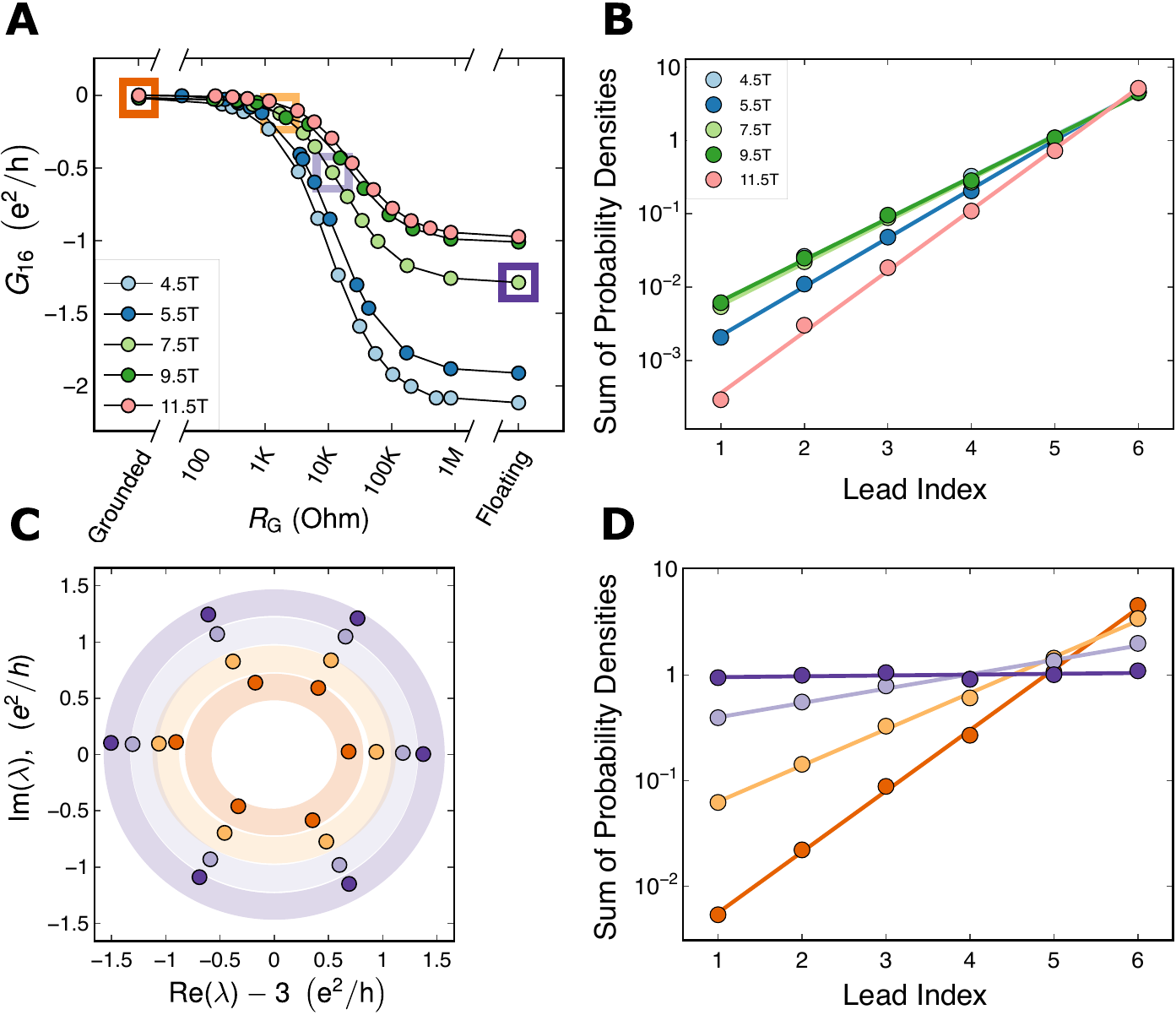}
	\captionlistentry{}
	\label{fig:SI:SPD_other_fields}

    \caption*{
	\noindent {\bf Fig. S7.} \textbf{Non-Hermitian skin effect. } 
	\textbf{(A)} Coupling term $G_{16}$ versus resistance $R_{\mathrm{G}}$ in logarithmic scale for various external magnetic fields. 
	The data points that are chosen for further graphical analysis in panels C and D are marked using squares.
	\textbf{(B)} SPD in logarithmic scale for the OBC setup at various values of the external magnetic field. 
	\textbf{(C)} Eigenspectra for the four selected points in units of $e^2/h$. The average diagonal element of the matrix is subtracted in order to center the plot around zero for better readability. The external magnetic field is $7.5$~T.
	\textbf{(D)}  Sum of probability densities (SPD) in logarithmic scale calculated from the measured $G=R^{-1}$ matrix for the highlighted data points as a function of the lead index, indicating a non-Hermitian skin effect. 
	The external magnetic field is $7.5$~T.	\hfill\break
	}
\end{figure}

\fig{fig:SI:SPD_other_fields} illustrates PBC to OBC tuning at different magnetic fields. 
The field values were chosen at the QHE plateaus ($\nu=2$ at 5.5~T and $\nu=1$ at 11.5~T) and between them ($4.5$~T and $7.5$~T, $9.5$~T). 
Panel A shows the dependence of the coupling term of the conductance matrix on the variable resistance $R_{\text{G}}$ (the experiment is performed in the same way as the one illustrated in the \fig{Fig. 2} in the main text). 
Panel B shows the SPD for the OBC case at the same magnetic fields. 
The skin effect localization is an exponential function of the lead index for magnetic fields corresponding to QHE plateaus, as well as between plateaus (the linear fit is indicated by the plain lines). 
Only the slope seems to be sensitive to the exact position of the Fermi energy with respect to the Landau levels.

Panels C and D show the eigenspectra and the SPD for a magnetic field between two QHE plateaus ($B=7.5$~T) for different values of the coupling term. 
The spectrum is slightly more elliptical than for a magnetic field corresponding to a QHE plateau, as shown in the \fig{Fig. 2} of the main text.
Nevertheless, the spectrum still shrinks when increasing the coupling term. 
Likewise, the skin effect and its dependence on the coupling term are very similar to the behavior observed at the QHE plateaus. 

\begin{figure}[tbh]
	\centering
	\includegraphics[width=.5\textwidth]{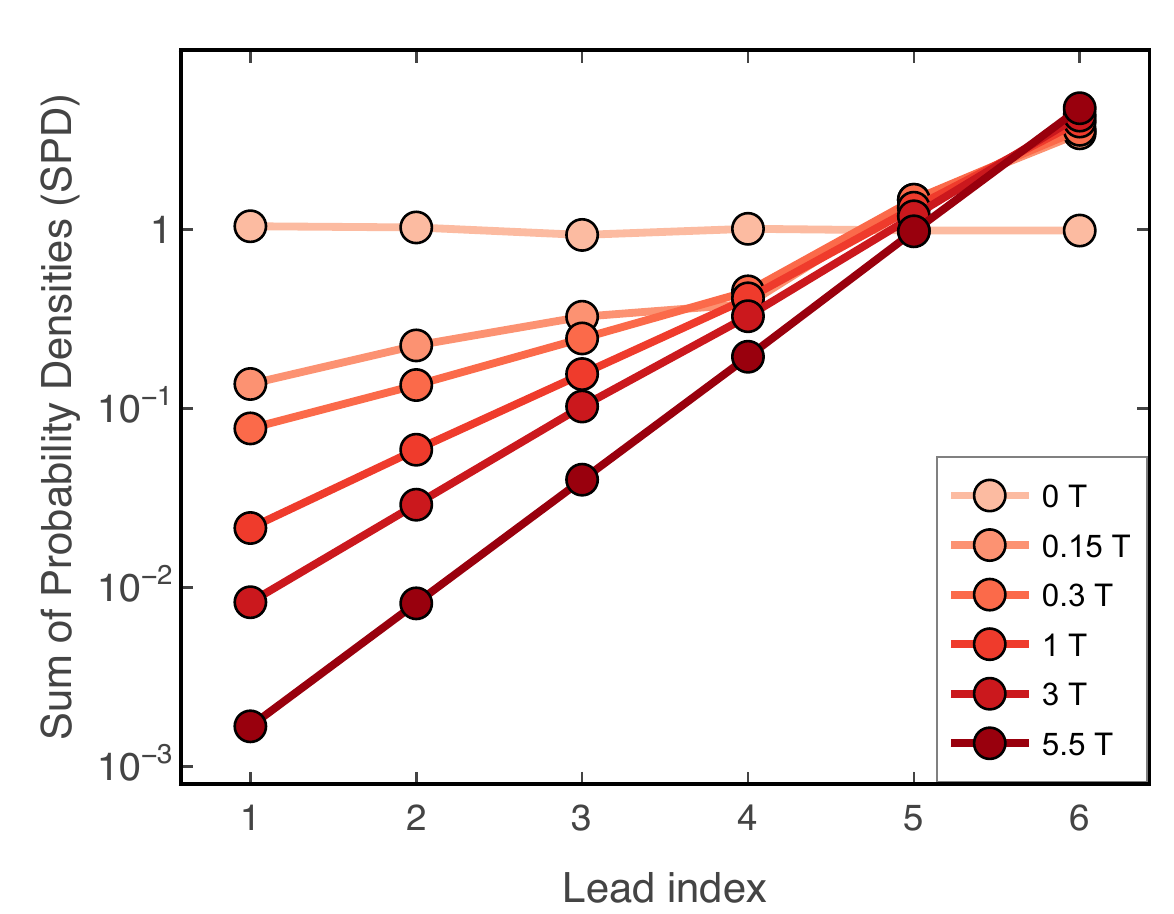}
	\captionlistentry{}
	\label{fig:SI:SPD_low_fields}

    \caption*{
    \noindent {\bf Fig. S8.} \textbf{Emergence of the non-Hermitian skin effect with increasing magnetic field.} 
    A decreasing SPD can be measured already at very weak magnetic field, but the exponential localization is observed only for larger field values.
    \hfill\break
    }
\end{figure}

\begin{figure}[tbh]
	\centering
	\includegraphics[width=0.8\textwidth]{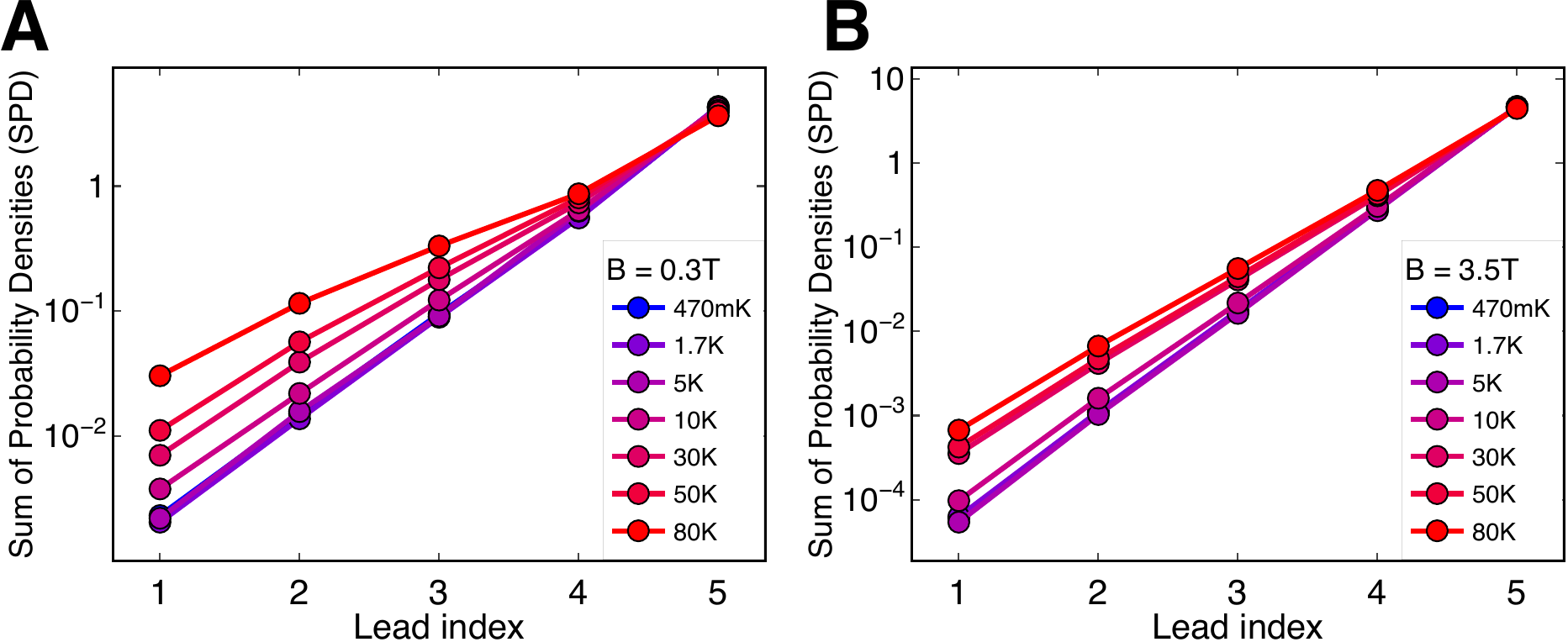}
	\captionlistentry{}
	\label{fig:SI:SPD_temp_comparison}
    
    \caption*{
	\noindent {\bf Fig. S9.} \textbf{Comparing the SPD for different temperatures} 
	\textbf{(A)} $B = 0.3$~T. Exponential localization of the states is stronger for lower temperatures. \textbf{(B)} $B = 3.5$~T, corresponding to the $\nu = 4$~QHE plateau.\hfill\break}
\end{figure}

\fig{fig:SI:SPD_low_fields} indicates how the non-Hermitian skin effect in OBC emerges when the magnetic field is ramped up.
At very weak magnetic fields, significant departures from an exponential localization are observed for the first sites of the chain, and the real-space invariant deviates strongly from the quantized value. 
This can be understood considering the effect of additional coupling terms in the matrix discussed in the Section \ref{SI:Additional_Coupling}. Similar deviation from the strong exponential localization of the skin effect is observed when the temperature is increased (see Fig. \ref{fig:SI:SPD_temp_comparison}), and is also attributed to the introduction of additional coupling terms in the $G$ matrix.

\clearpage
\newpage

\subsection{Measurements at different temperatures and charge density} \label{SI:DiffTempAndDensity}

\begin{figure}[tbh]
	\centering
	\includegraphics[width=0.9\textwidth]{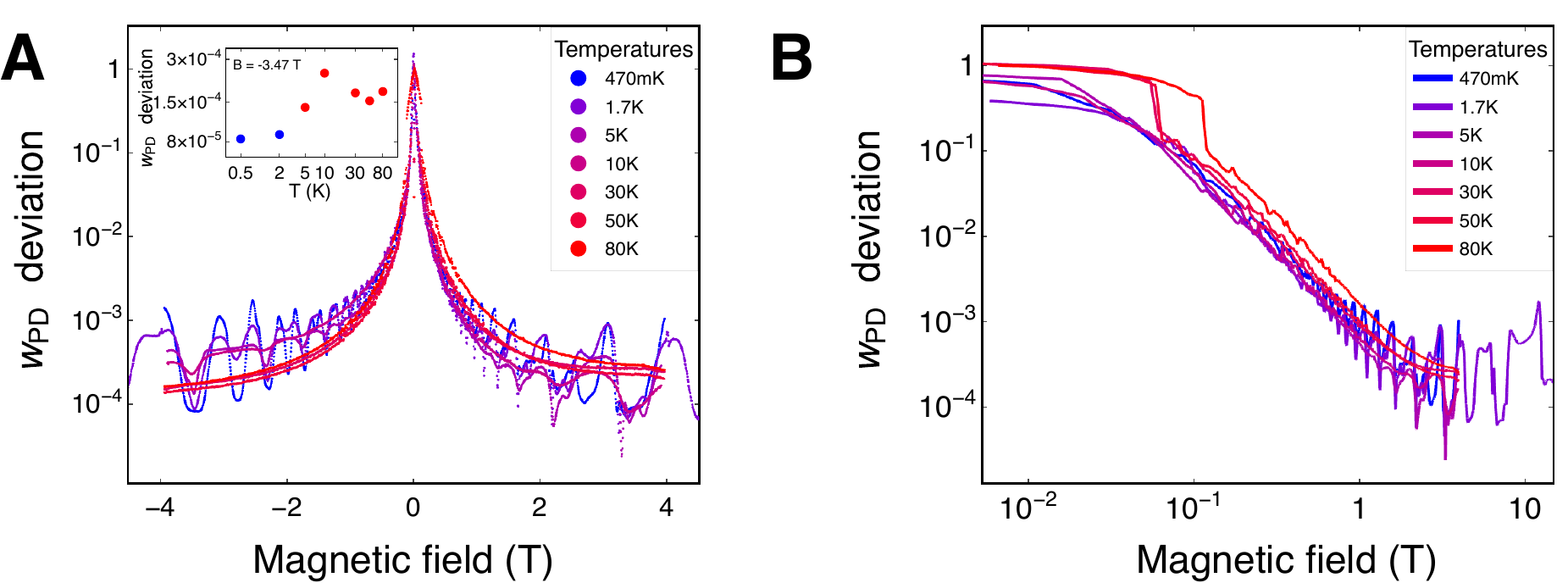}
	\captionlistentry{}
	\label{w_PD-Temp}
    
    \caption*{
	\noindent {\bf Fig. S10.}
    \textbf{Magnetic field dependence of the invariant quantization at different temperature} 
    (A) Deviation from the quantization of the topological invariant $w_{PD}$ at different temperatures between 470~mK and 80K and between $B=-4$~T and $B=4$~T. Measurements are done in the 5 sites OBC configuration. Similarly to the \fig{Fig. 3}~(C), the inset shows the invariant deviation measured at various temperatures in the QHE plateau at $B = -3.47$~T. The blue color denotes temperatures for which a quantum Hall plateau is measured whereas the red color denotes temperatures for which only dips (and no plateaus) are measured. (B) Same measurement plotted in a logarithmic-logarithmic scale.
    \hfill\break
	}
\end{figure}

\begin{figure}[tbh]
	\centering
	\includegraphics[width=0.45\textwidth]{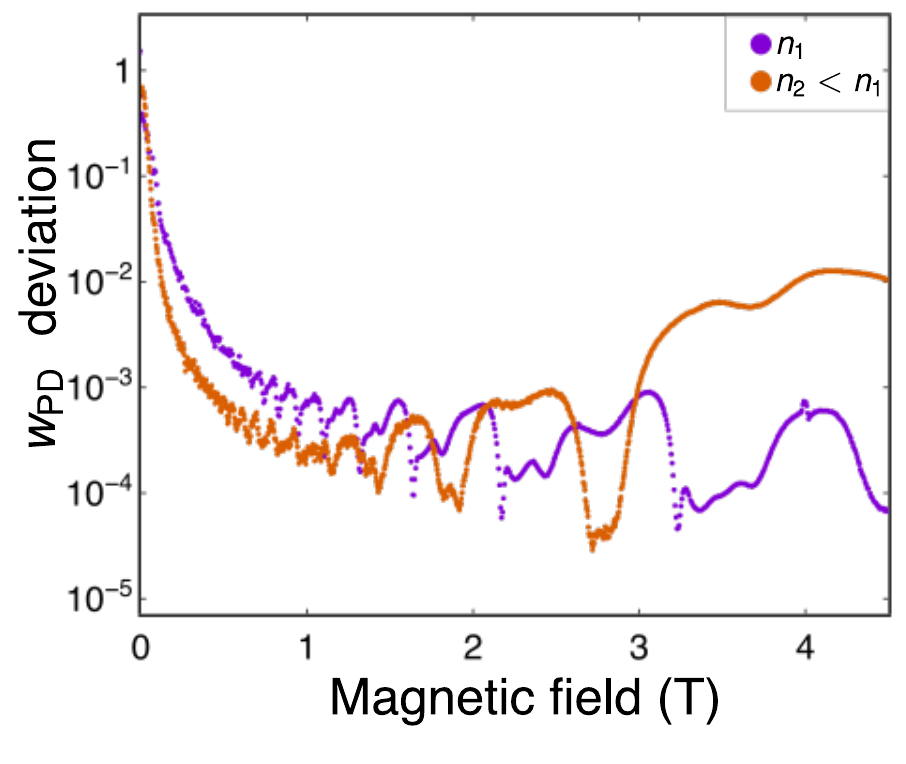}
	\captionlistentry{}
	\label{w_PD-Density}
    \caption*{
	\noindent {\bf Fig. S11.} \textbf{Invariant quantization for different charge density} 
    Magnetic field dependence of the invariant quantization in the 5 sites configuration and for two different cooldowns corresponding to electronic densities that differ by 20\%. \hfill\break
	}
\end{figure}

In order to test deviations from Eq.~\eqref{eq:Kirchhoff} in the main text, and to test the robustness of our invariant 
with respect to the type of transport
regime (ballistic versus diffusive), we did measurements at temperatures ranging between 470~mK and 80~K. The results for the skin effect are presented above in Section \ref{SI:OBC_to_PBC_other_fields}, and we focus here on the topological invariant $w_{PD}$. When zooming
in
on the field dependence of the quantization at low magnetic field, the deviation from $\pm 1$ is almost temperature independent (within the fluctuations induced by the asymmetry), as seen in the \fig{w_PD-Temp}~(A),  and vanishes quadratically (\fig{w_PD-Temp}~(B)). At low temperatures and large magnetic fields ($B \sim 4$~T), the quantization drops down to a minimum when the Fermi energy is set between two Landau levels due to the reduction of the backscattering terms already mentioned above. As in \fig{Fig. 3}~(C) in the main text, the inset in \fig{w_PD-Temp}~(A) shows that this is true also for negative magnetic field and despite the asymmetry induced by the onsite disorder as explained in \ref{SI:JumpsAndAsymmetry}. This highlights the importance of  being in such a quantum state in order to match the case of maximized non-reciprocical Hatano-Nelson chain.

We did two sets of measurements with two different cool-downs, leading to a significant difference in the charge density of about 20\% between these measurements. Contrary to the effect of the temperature, the density appears to have an influence on the decay of the deviation from quantization of the  $w_{PD}$ invariant, the lower density showing a sharper decay with similar saturation quantization (\fig{w_PD-Density}).

\subsection{Invariant for different system configurations} \label{SI:invariants_for_different_leads}

\begin{figure}[h]
	\centering
	\includegraphics[width=.4\textwidth]{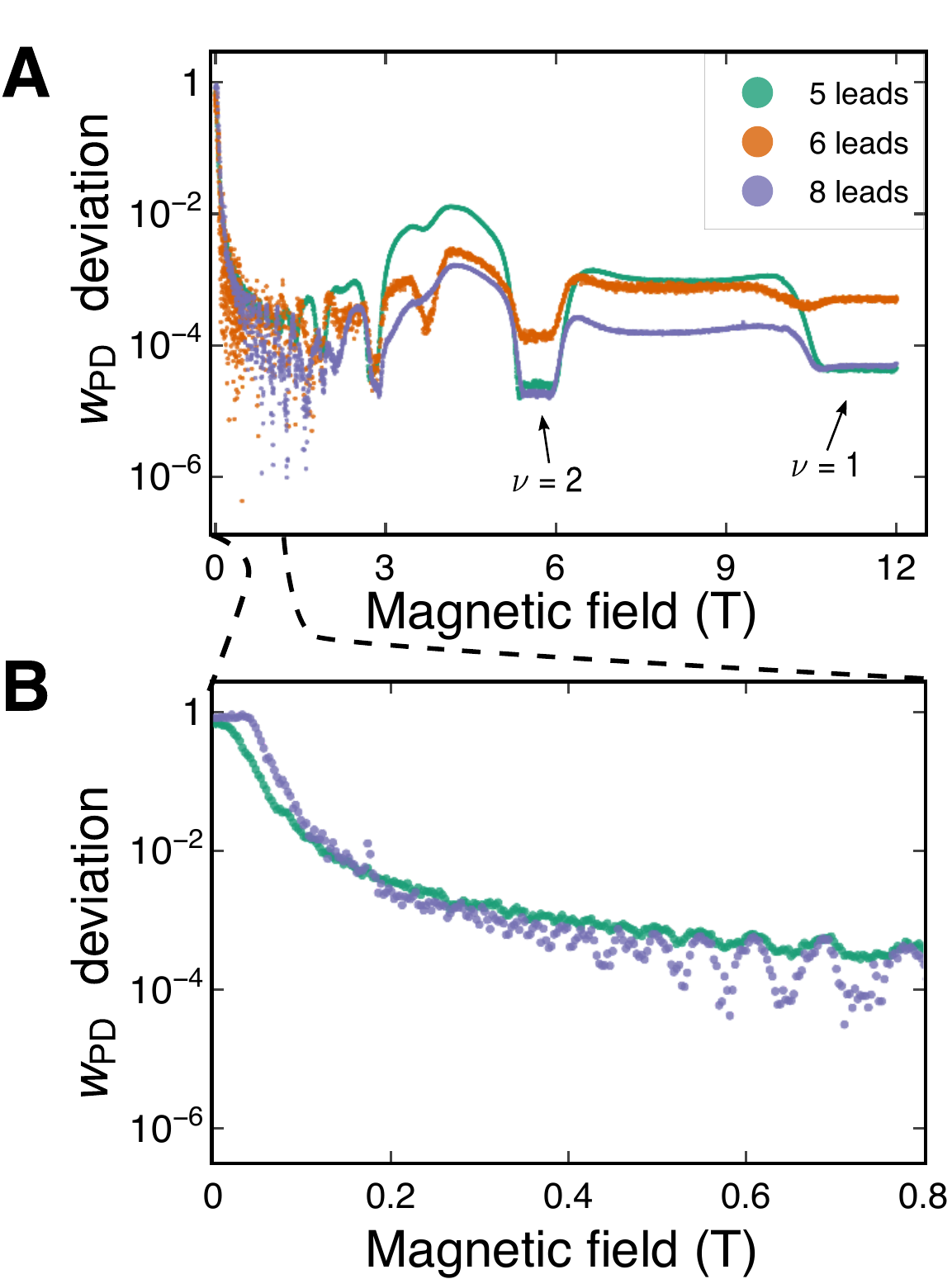}
	\captionlistentry{}
	\label{fig:SI:additional_invariants}

    \caption*{
	\noindent {\bf Fig. S12. Invariant quantization for different system configurations.}
	\textbf{(A)} Deviation of $w_{\rm PD}$ from the quantized value $-1$, showing the stability of the skin effect over a large magnetic field range.
	The deviation is plotted in logarithmic scale versus magnetic field for the 5, 6, and 8 site setup. The different grounding of the chain are realized by four contacts in the 5 leads configuration and by a single contact in the 6 and 8 leads configuration.
	\textbf{(B)} Zoom in of panel A at low magnetic fields for the 5 and 8 site setup, showing the lowest noise level.\hfill\break
	}
\end{figure}

The magnetic field dependence of the conductance matrix was measured for the 5, 6, and 8 site configurations. 
The magnetic field dependence of the real-space invariant is then calculated for each configuration, and the comparison of the invariant quantization for different number of sites is shown in \fig{fig:SI:additional_invariants}. 

As shown in panel A, the invariant quantization always improves when the Fermi energy is between two Landau levels.
These magnetic fields coincide with the minima of the longitudinal magnetoresistance, both in the Shubnikov-de Haas oscillations regime and in the QHE regime. 
It corresponds to a situation where the backscattering through the bulk states is significantly reduced. 
The best quantization is achieved for $\nu=2$ ($B\sim 5.6$~T) or for $\nu=4$ ($B\sim 3$~T).

We focus first on the situation where the Fermi energy lies between two Landau levels.
For the 
5-site
configuration, we used four grounded contacts to disconnect the first site from the last 
site,
whereas only a single ground was used for the 
6- and 8-site
chain. The comparison between the 
latter two
configurations indicates that finite-size distortions of the non-Hermitian skin effect are 
non-negligible
for 6 sites. The quantization of the invariant is thus limited by the number of sites of the chain. For both the 
5-site and 8-site
configurations, the invariant quantization is the same at $\nu=1$, $\nu=2$ and $\nu=4$, indicating that, deep in the quantum Hall regime,
the finite-size effects for a 5-site system are fully compensated by the multiple grounds used to disconnect the last and first site.

On the contrary, when the Fermi energy is positioned at a Landau level, inducing possible backscattering through bulk states, and therefore additional coupling terms in the $G$ matrix (see \ref{SI:Additional_Coupling}), the 5-site configuration shows a significantly lower quantization than the 8-site configuration and even lower than the 
6-site
configuration.
This suggests that the role played by finite-size effects becomes more important than the grounding configuration at the plateau transitions. For such magnetic fields, the quantization could be improved by increasing the number of leads. 
The dependence of the quantization level on the number of leads is further confirmed at low magnetic fields, where the 5-site system shows, again, a less quantized invariant compared to the 8-site system.

\subsection{Jumps and field asymmetry of the invariant} \label{SI:JumpsAndAsymmetry}
If the conductance matrix were a perfectly homogeneous Toeplitz  matrix, the invariant would have the property
\begin{equation}
    w_{PD}(G) = -w_{PD}(G^\dag).
\end{equation}
Further, in ideal conditions the conductance matrix is real and follows
\begin{equation}
    G(B) = G^T(-B).
\end{equation}
As a consequence, the quantization of the invariant shown in Fig.~\ref{w_PD-Temp} should be symmetric for positive and negative magnetic fields.
The reason for the observed asymmetry comes from the fact that the conductance matrices are indeed not perfect Toeplitz matrices.
The chain is not perfectly mirror symmetric and disorder prefers one side to the other.
This comes from geometric effects related to where the contacts are positioned 
relative
to the ground contact, leading to slightly larger elements towards one end of the chain.
This has an effect on the quality of the invariant quantization, albeit very small and only visible as a deviation from the quantized value by less than $10^{-3}$ due to the low noise level.

Furthermore, we see in Fig.~\ref{w_PD-Temp} that at higher temperatures 
and
lower magnetic fields
($T \ge 50K$ and $B \approx 100 ~mT$)
the invariant changes discontinuously. This discontinuity can be understood in terms of long-range hopping between contacts.
For systems with short-range hoppings and open boundary condition, such smoothly varying system parameters will only cause smooth changes in the invariant quantization.
However, at higher temperatures, longer range hoppings through bulk channels become more important and cause deviation from the OBC case, since now conductances are allowed between distant leads.
By smoothly changing the magnetic field, different longer range left and right hopping compete with each other. When 
two such
hopping parameters cross, we can expect discontinuities in
the polar decomposition, which leads to jumps in the invariant.

\subsection{Details of the iterative measurement} \label{SI:iteration_explanation}

The iteration process at the $j^\mathrm{th}$ iteration consists in injecting a sine-wave current vector $\mathbf{I}(j,n)$ where $n$ stands for an effective time of the sine-wave functions. Each component $i$ of the current vector is the addition of a constant $C$ independent of $i$ (see below) and $n$ and a sine function characterized by its phase  $\phi_{i}(j)$, and of 
its amplitude $A_{i}(j)>0$. The component $i$ of the current vector corresponds to the current injected into the lead $i$. 
$A_{i}(j)$ and $\phi_{i}(j)$ are randomly generated for $j=0$. In order to generate the sine functions, we discretize the signal in $N$ regularly spaced points. Each point indexed by $n$ with $n \in \{ 0, \ldots, N-1 \}$ coincides with a $2\pi n / N$ phase in the sine functions. 

For the six-site configuration we used, the current vector reads
\begin{equation}
\mathbf{I}(j,n)=
\left[\begin{array}{l}
I_1(j,n)\\
I_2(j,n)\\
I_3(j,n)\\
I_4(j,n)\\
I_5(j,n)\\
I_6(j,n)
\end{array}\right]
=
\left[\begin{array}{l}
A_{1}(j) \sin \left[2\pi n/N+\phi_{1}(j)\right] + C\\
A_{2}(j) \sin \left[2\pi n/N+\phi_{2}(j)\right] + C\\
A_{3}(j) \sin \left[2\pi n/N+\phi_{3}(j)\right] + C\\
A_{4}(j) \sin \left[2\pi n/N+\phi_{4}(j)\right] + C\\
A_{5}(j) \sin \left[2\pi n/N+\phi_{5}(j)\right] + C\\
A_{6}(j) \sin \left[2\pi n/N+\phi_{6}(j)\right] + C
\end{array}\right] \times \sin{(\omega t)},
\label{SI:eq:iteration_init}
\end{equation}
The amplitudes are normalized in our experiment such that $\max_{i} (A_{i}) + C =150$~nA, and we choose $C=75$~nA such that all the components of the current vector are positive for any value of $n$. 
We used the lock-in amplifier's voltage source and a polarization resistance of 1~M$\Omega$ to generate AC current sources, whose amplitudes correspond to the related component of the current vector. 
We used a total number of $N=30$ points, which allows for a reliable determination of the amplitudes and phases of the current vector's sine-wave components.

For a 6-site setup and for the $j^\textrm{th}$ iteration, a current vector as defined in Eq.~\eqref{SI:eq:iteration_init} is applied 30 times with $n=0,...,29$, and the corresponding voltage vector is measured for each $n$. 
This allows us to determine the relative amplitude and the phase of the different components of the voltage vector for the $j^\textrm{th}$ iteration. 
The amplitude and phase of the current vector $j+1$ is given by the renormalization of the voltage vector $j$ such that $\max_{i} (A_{i}) = C =75$~nA.

\begin{figure}[tbh]
	\centering
	\includegraphics[width=.7\textwidth]{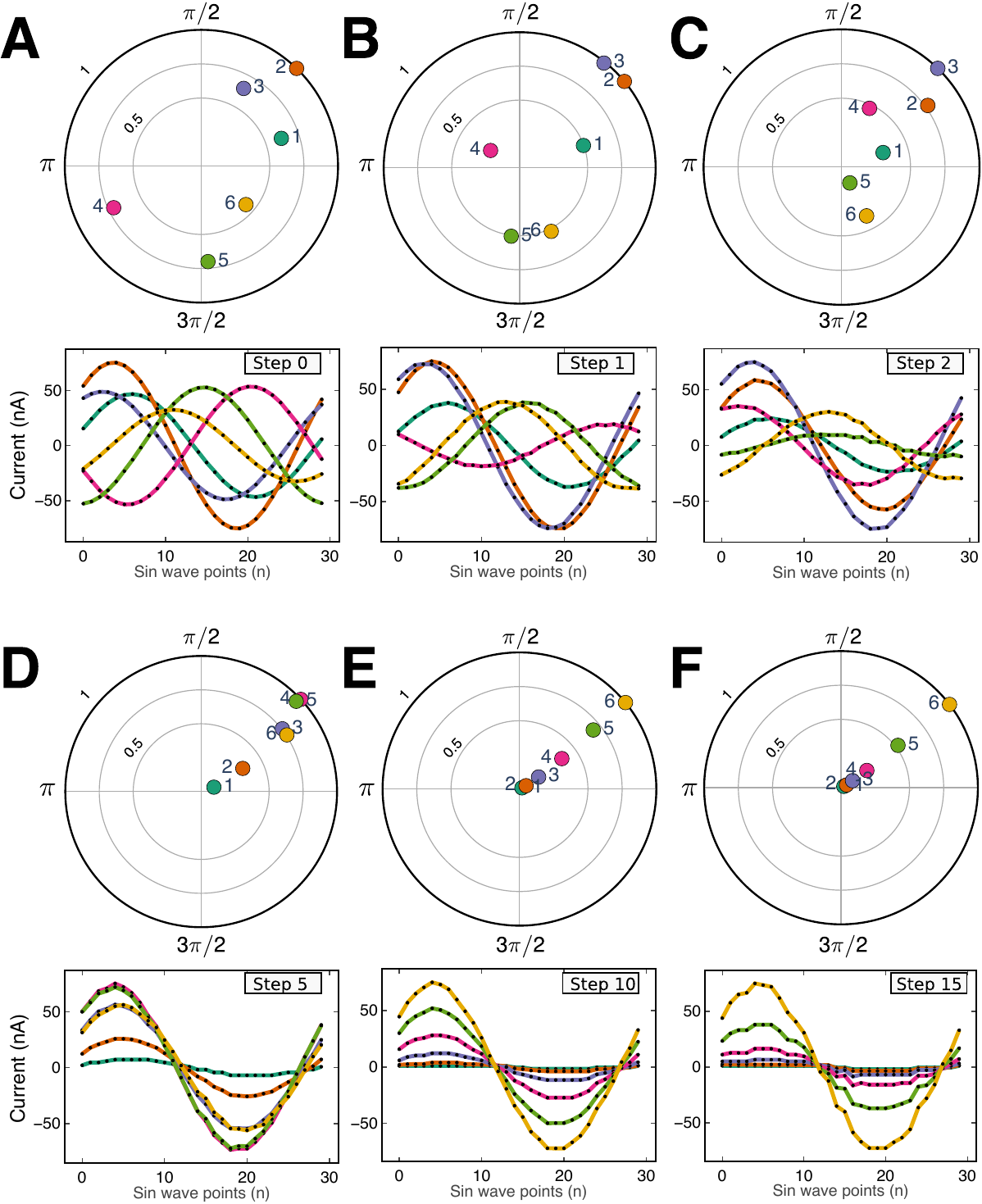}
	\captionlistentry{}
	\label{fig:SI:generated_sine_waves}

    \caption*{
    \noindent {\bf Fig. S13.} \textbf{Generated sine waves.} 
    \textbf{(A)} Randomly generated initial currents displayed on a unitary circle in polar coordinates for a 6-site OBC set-up. 
    The radius and the angle correspond respectively to the amplitude $A_{i}(0)$ and the phase $\phi_{i}(0)$ in Eq.~\eqref{SI:eq:iteration_init}. The lower panel shows the injected signal representing the different components of the vector for $n=0$.
    \textbf{(B-F)} Evolution of the signal throughout the steps of the iteration process.\hfill\break
    }
\end{figure}

The evolution of the measured voltage vector components versus iteration index is shown as an animation \cite{zenodocode} for the 6-site setup, with snapshots shown in Fig.~\ref{fig:SI:generated_sine_waves}.

\begin{figure}[tbh]
	\centering
	\includegraphics[width=.6\textwidth]{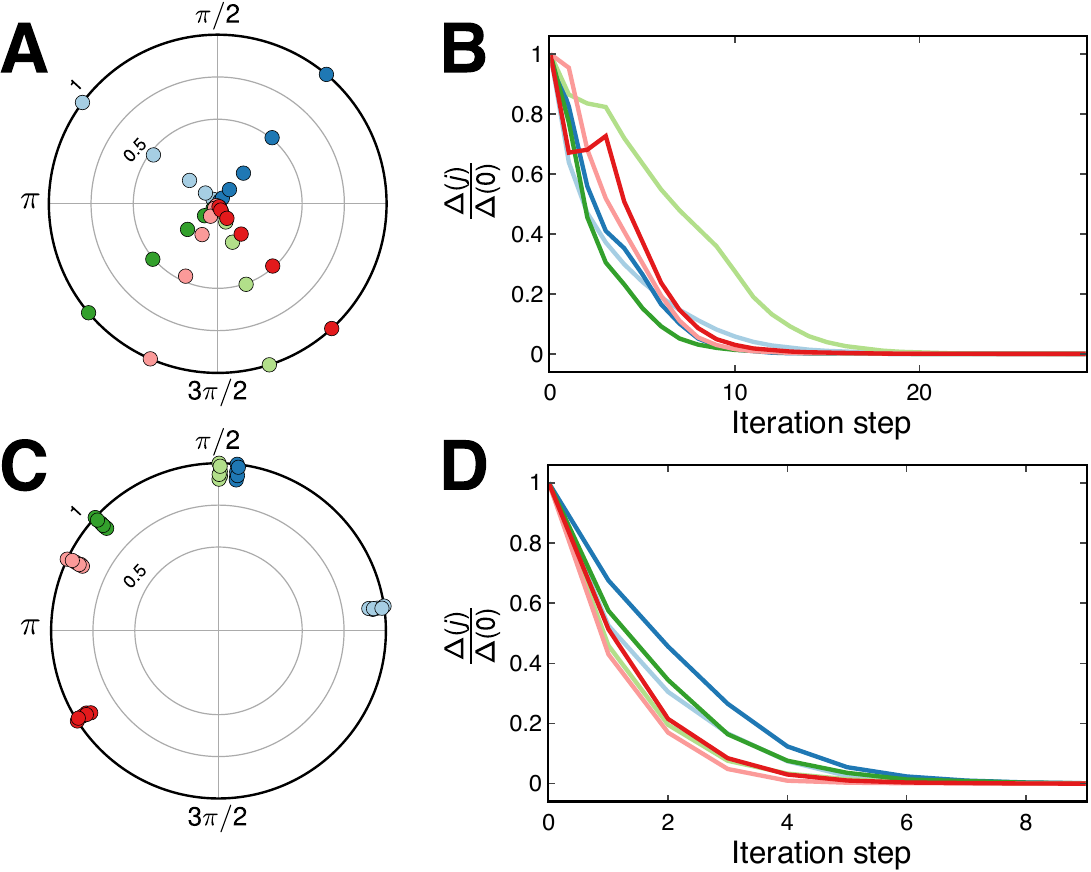}
	\captionlistentry{}
	\label{fig:SI:all_iterations}

    \caption*{
    \noindent \noindent {\bf Fig. S14} \textbf{All experimental power iteration runs.} 
    \textbf{(A)} Different final current vectors of the system, corresponding to different initial random vectors, displayed on a unitary circle in polar coordinates for a 6-site OBC set-up. 
    The radius and the angle correspond respectively to the amplitude $A_{i}(0)$ and the phase $\phi_{i}(0)$ in Eq.~\eqref{SI:eq:iteration_init}. 
    Different colors represent different initial current vectors. The final current configurations are always in phase (regarding the different components of the vector) and always show a non-Hermitian skin effect.
	\textbf{(B)} 
	 Plot of the standard deviation of the current vector as defined in Eq.~\eqref{SI:eq:StandardDeviation} as a function of the iteration index for OBC \textbf{(C)} Elements of the final currents injected into the system for a 6-site PBC set-up.
	\textbf{(D)}  Plot of the standard deviation of the current vector as defined in Eq.~\ref{SI:eq:StandardDeviation} as a function of the iteration index for PBC.
	\hfill\break
	}
\end{figure}

We have measured 8 different iteration processes in the OBC, all of them ending in a final current configuration corresponding to the eigenvector associated to the same eigenvalue of the $G$-matrix, regardless of the starting configuration. 
This eigenvector has all components in-phase. Its related eigenvalue is real, meaning that current and voltage of each contact are in-phase. The final current and voltage configuration thus presents a skin effect (\fig{fig:SI:all_iterations}A). 
We have also performed 6 iteration processes in the PBC configuration (\fig{fig:SI:all_iterations}C), to confirm that the skin effect is not present in this case.

In order to quantify how fast the iteration process converges on its final current configuration, we introduce the quantity $\Delta$ that is a function of the iteration step $j$, and that is defined as 
\begin{equation}
\Delta (j) = \sum_{{n,i}} \left( I_{i}(j,n) - I_{i}(\infty,n)\right)^2,
\label{SI:eq:StandardDeviation}    
\end{equation}
where $\mathbf{I}(j,n)$ is the current vector defined above, and where $j \rightarrow \infty$ corresponds to the final iteration step. \fig{fig:SI:all_iterations}B and \fig{fig:SI:all_iterations}D represent the value of $\Delta (j) / \Delta (0)$ for OBC and PBC respectively. 
We observe a convergence to the final configuration in about 10 iterations for OBC, which is significantly reduced for the 
PBC, where only about 5 iterations are needed
to reach the final configuration.

To understand the convergence of the system gradually to an eigenvector of the $G$ matrix, one needs to consider the eigenvalues $\lambda_{i}$ and eigenvectors $\mathbf{u}_{i}$ of the $G$ matrix. 
Such a convergence can be simply understood considering the projection of the initial current vector $\mathbf{I}(0)$ onto the eigenvector $\mathbf{u}_{\rm min}$ associated to the eigenvalue $\lambda_\text{min}$ with $|\lambda_{\rm min}|=\min_i(|\lambda_{i}|)$ in absolute value. 
We have then
\begin{equation}\label{eq:V_j}
	\mathbf{I}(j) \propto R^j \mathbf{I}(0) = G^{-j} \mathbf{I}(0) \xrightarrow{j \rightarrow +\infty} \left(\lambda_{\rm min}\right)^{-j} \left(\mathbf{I}(0) \cdot \mathbf{u}_{\rm min} \right)\mathbf{u}_{\rm min}.
\end{equation}

Since the eigenvectors of the $G$ matrix all show a skin effect, the system always converges to a current and voltage configuration that shows this skin effect. 
This convergence is a direct signature of the topologically non-trivial non-Hermiticity of the system.

\end{document}